\documentclass[graybox]{svmult}

\usepackage{type1cm}        % activate if the above 3 fonts are
\usepackage{makeidx}         % allows index generation
\usepackage{graphicx}        % standard LaTeX graphics tool
\usepackage{multicol}        % used for the two-column index
\usepackage[bottom]{footmisc}% places footnotes at page bottom

\usepackage{newtxtext}       % 
\usepackage{newtxmath}       % selects Times Roman as basic font

\makeindex             % used for the subject index
\begin{document}

\title*{Low-mass and sub-stellar eclipsing binaries in stellar clusters}
\author{Nicolas Lodieu$^{1,2}$, Ernst Paunzen$^{3}$, and Miloslav Zejda$^{3}$}
\institute{$^{1}$ Instituto de Astrof\'isica de Canarias (IAC), Calle V\'ia L\'actea s/n, E-38200 La Laguna, Tenerife, Spain, \email{nlodieu@iac.es} \\
\and 
$^{2}$ Departamento de Astrof\'isica, Universidad de La Laguna (ULL), E-38206 La Laguna, Tenerife, Spain \\
\and
$^{3}$ Department of Theoretical Physics and Astrophysics, Masaryk University, Kotl\'arsk\'a 2, 611 37 Brno, Czech Republic,
\email{epaunzen@physics.muni.cz, zejda@physics.muni.cz}}
\maketitle

\abstract{
%Each chapter should be preceded by an abstract (no more than 200 words) that summarizes the content. 
%The abstract will appear \textit{online} at \url{www.SpringerLink.com} and be available with unrestricted access. 
%This allows unregistered users to read the abstract as a teaser for the complete chapter.\newline\indent}
{\bf{(no more than 200 words)}}. \newline
We highlight the importance of eclipsing double-line binaries in our understanding on 
star formation and evolution. We review the recent discoveries of low-mass and sub-stellar eclipsing binaries belonging to star-forming regions, open clusters, and globular 
clusters identified by ground-based surveys and space missions with high-resolution spectroscopic follow-up. These discoveries provide benchmark systems with known 
distances, metallicities, and ages to calibrate masses and radii predicted by state-of-the-art evolutionary models to a few percent. We report their density and 
discuss current limitations on the accuracy of the physical parameters. We discuss future opportunities and highlight future guidelines to fill gaps in age and metallicity to improve further our knowledge of low-mass stars and brown dwarfs.
}

%
%%%%%%%%%%%%%%%%%%%%%%%%%%%%%%%%%%%%%%%%%%%%%%%%
%%%%%
%%%%%%%%%%%%%%%%%%%%%%%%%%%%%%%%%%%%%%%%%%%%%%%%
%
\section{The importance of eclipsing binaries}
\label{book_EBs:EBs_important}
\subsection{Scientific context}
\label{book_EBs:EBs_context}

Low-mass M dwarfs are the most common stars in the Solar neighbourhood and, more
generally, in the Universe, accounting for about 70\% of the entire population of hydrogen-burning
stars with masses below 0.6 M$_{\odot}$ \cite{henry97,winters15}. Determining their physical parameters 
(luminosity, mass, radius) is fundamental to understand stellar evolution and constrain theoretical isochrones. 
At lower temperatures, the numbers of brown dwarfs, objects unable to fuse hydrogen in their 
interiors \cite{kumar63a,tarter76,chabrier00a}, with accurate mass and radius measurements 
remain very limited.

Eclipsing binaries (EBs) are systems with two components lying on the same plane with respect
to the observer transiting each other periodically.
They are fundamental probes of stellar evolution and stellar candles to measure distances of 
clusters \cite{southworth05} because the radius and mass of each component can be derived
with high precision from the photometric light-curves and radial velocity monitoring, respectively 
\cite{torres10a}.

The numbers of EBs has increased dramatically over the past two decades thanks to the advent
of large-scale photometric and spectroscopic surveys as well as space missions.
Following up on the original reviews on fundamental parameters of stars derived from EBs
\cite{popper80a,andersen91a}, a catalogue of detached EBs with their main physical parameters 
including masses and radii determined to
precisions better than a few percent is constantly updated \cite{southworth14a}\footnote{the
catalogue is maintained at http://www.astro.keele.ac.uk/jkt/debcat/}. Another public databases 
with physical parameters of EBs
is the Binary Star Database \cite{svechnikov04a}\footnote{the Binary Star Database
can be found at http://bdb.inasan.ru/}, which contains physical and positional parameters of the 
components of 120,000 stellar multiple systems compiled from a variety of published catalogues 
and databases. 

Other unrelated projects contributed, currently supply, and will add to our knowledge of EBs. 
As a few example, the OGLE project principally devoted to microlensing provides huge amount 
(several hundred thousands) of eclipsing systems over a wide range of mass and evolutionary states
towards the Galactic Bulge \cite{soszynski17a}\footnote{http://ogle.astrouw.edu.pl/}.
The All Sky Automated Survey (ASAS)\footnote{http://www.astrouw.edu.pl/asas/} is a low cost 
project dedicated to constant photometric monitoring of the full sky to study variable phenomenon 
of any kind, including the study of EBs \cite{helminiak12a}.
The Large Sky Area Multi-Object Fiber Spectroscopic Telescope (LAMOST) is a large Chinese
project dedicated to spectroscopy of several millions of stars with spectral classification.
This program has brought several thousands of EBs over a wide range of spectral type
\cite{zhang18a}, with some masses and radii determined for a few low-mass systems \cite{lee17a}.
Other programs mainly dedicated to the discovery and tracking of minor bodies, such as the 
Catalina Sky Survey \cite{drake09a,mahabal11,djorgovski12}\footnote{https://catalina.lpl.arizona.edu/}
or the Asteroid Terrestrial-impact Last Alert System (ATLAS) project \cite{tonry18a}\footnote{https://atlas.fallingstar.com/}
do regularly contribute to the discovery of EBs.

This review focuses on low-mass EBs with at least one M dwarf or sub-stellar companions
members of star-forming regions (Section \ref{book_EBs:EBs_SFRs}), open clusters
(Section \ref{book_EBs:EBs_OCs}), and globular clusters (Section \ref{book_EBs:EBs_GCs}). 
We discuss the frequency of EBs in clusters and the impact of age, metallicity, and stellar
variability/activity on their physical parameters (Section \ref{book_EBs:discussion}).
This review is timely due to the most recent contribution of the Kepler K2 mission 
\cite{borucki10,howell14} to our knowledge of low-mass EBs in clusters, whose masses and 
radii can directly be confronted to model predictions.
The study of EBs requires huge observing time investment on both photometric and spectroscopic 
sides needed to infer masses and radii, as demonstrated by the WIYN cluster survey 
\cite{geller09,milliman14,leiner15}, the Young Exoplanet Transit Initiative (YETI) focusing
on young clusters \cite{neuhaueser11}, the Palomar Transient Factory survey 
\cite{rau09,law09}\footnote{https://www.ptf.caltech.edu/iptf}, and the Kepler K2 mission.
We finish this review with a list of requirements and prospects to fill in gaps in the 
Hertzsprung-Russell (H-R) diagram (Section \ref{book_EBs:EBs_future}).

\subsection{How masses and radii are determined observationally}
\label{book_EBs:EBs_determination}
The first attempts to determine the parameters of eclipsing binaries and their components were done in the end of 19$^{th}$ century. Up to late 1960's and 1970's year a series of method were developed and used on light curves and radial velocity curves series to subtract and interpolate data from tables of different quantities (more details in \cite{RussellMerrill1952} or \cite{Zverev1947}). The spread of computers fasten the development of many codes such as EBOP (Eclipsing Binary Orbit Program) \cite{Etzel1981}, SEBM (Standard Eclipsing Binary Star Model) \cite{Budding1973,BuddingZeilik1987}, WINK \cite{wood1971,wood1972,wood1973}, LIGHT2 \cite{hill1979,HillRucinski1993}, version of WUMa \cite{rucinski1973,rucinski1974} and others.

In 1971, Wilson and Devinney published the results of their code (hereafter WD) where they used for the first time the least-squares method to extract the parameters of light curves \cite{WD1971,WD1972,WD2008,WD2014}. This WD code has been regularly upgraded up to now and could be downloaded from author's ftp\footnote{ftp://ftp.astro.ufl.edu/pub/wilson/}. Independently, users created graphical user interfaces and some minor upgrades. However, the project PHOEBE \cite{PrsaZwitter2005} is not only GUI for calculations based on WD core, nowadays it has become a more general code to models both the photometric light curve and radial velocity curves of eclipsing binaries. The new version of PHOEBE2, which is still under development\footnote{http://phoebe-project.org}, contains more physics and improved mathematical methods for the solutions of eclipsing binaries \cite{Prsa2016,Horvat2018,Prsa2018}.

Independent codes like MECI (Method for Eclipsing Component Identification) and DEBil (Detached Eclipsing Binary Light curve fitter) \cite{devor2005,devor2006}, EBAS (Eclipsing Binary Automated Solver) \cite{tamuz2006a,tamuz2006b}, FOTEL \cite{hadrava2004}, JKTEBOP\footnote{http://www.astro.keele.ac.uk/~jkt/codes/jktebop.html}, ROCHE \cite{pribulla2012}, Nightfall\footnote{https://www.hs.uni-hamburg.de/DE/Ins/Per/Wichmann/Nightfall.html}, Binary Makers (BM) \cite{Bradstreet2005} are used in limited numbers of publications. Two authors of these codes also collect binary stars solutions -- David H. Bradstreet, author of Binary Maker, manages the Catalog and AtLas of Eclipsing Binaries (CALEB) based only on BM solutions\footnote{http://caleb.eastern.edu/} and John Southworth the DEBCat catalogue\footnote{http://www.astro.keele.ac.uk/~jkt/debcat/}, which contains physical properties of well-studied detached eclipsing binaries where errors on the mass and radius determinations are mostly below 2\%.

\begin{figure}
\sidecaption
\includegraphics[width=0.90\linewidth]{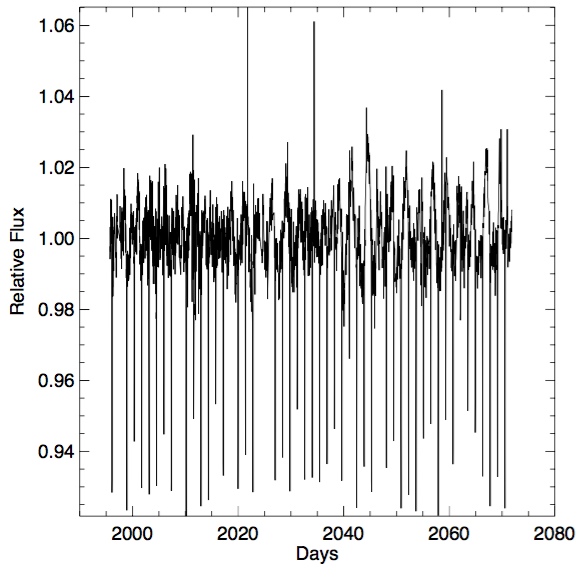}
\includegraphics[width=0.90\linewidth]{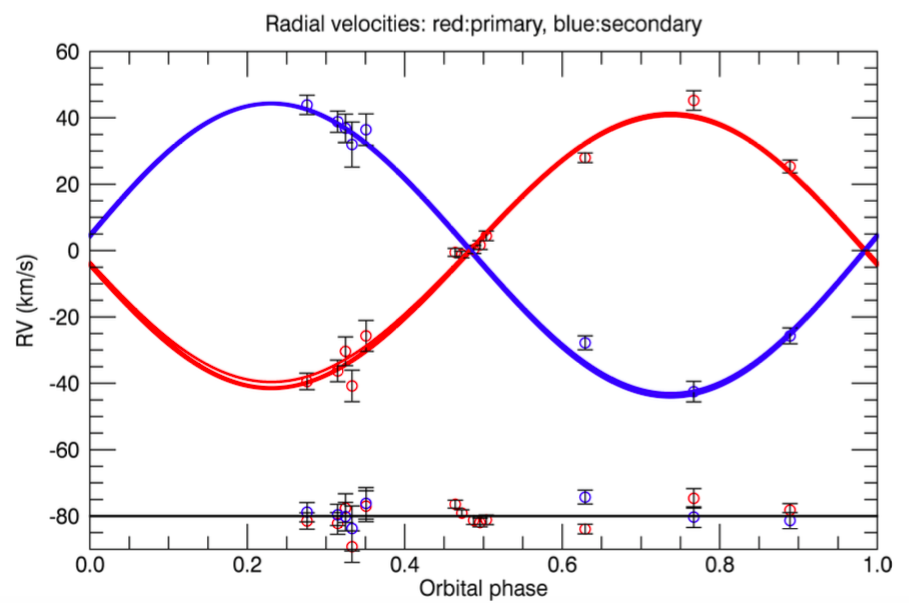}
\caption{
{\it{Top:}} Kepler light-curve for USco\,J16163068$-$2512201 (=\,EPIC\,203710387) 
over the full $\sim$76 days of the K2 campaign two in Upper Scorpius.
{\it{Bottom:}} Radial velocity measurements as a function of phase for the primary 
(blue symbols) and the secondary (red symbols) of USco\,J16163068$-$2512201\@.
Figure taken from \cite{lodieu15c}.
}
\label{fig_book_EBs:fig_LC_RV_EBs}
\end{figure}

From the aforementioned codes, we can estimated the physical parameters of each component of a multiple system. The main parameters derived from the analysis of the light curve(s) are orbital period, (possibly( eccentricity, orbital inclination, relative ratio of the radius of the primary and secondary of the system considering the separation of the components (top panel in Fig.\ \ref{fig_book_EBs:fig_LC_RV_EBs}), system luminosity and photometric mass ratios. However, in some cases photometric mass ratios might be unreliable in comparison to spectroscopic mass ratios \cite{terrell2005,hambalek2013}. 

The light curve solution usually requires photometric data in at least two filters. The availability of only one passband data means that some of parameters must be estimated and/or fixed. The effective temperature of primary is inferred from its spectral type or colour indices. Limb darkening coefficients are interpolated from tables e.g.\ \cite{vanhamme1993}, gravity brightening and bolometric albedo coefficients are set according to the expected type of stellar atmospheres. Then, except for the parameters mentioned above, one can determine the surface potentials, the rotational/orbital synchronicity, and the third light. 

The situation improves rapidly when radial velocity measurements are available for both components (Fig.\ \ref{fig_book_EBs:fig_LC_RV_EBs}). In this case, it becomes possible to figure out the spectroscopic mass ratio and distance of the components, which serve as a scaling factor for the radii of each component. The combination of photometric and spectroscopic datasets leads to the determination of absolute eclipsing binary parameters in physical units, including masses, sizes, and luminosities of both components as well as distance from Earth. In this process, we can also calculate the atmosphere model and corresponding parameters (see e.g. \cite{claret2011,claret2012,claret2013,claret2017,claret2018,neilson2013}).

Some of aforementioned codes coupling light curve and radial velocity solutions are also capable to process additional kinds of parameters like timings of minima, interferometric measurements, and so on.

%
%%%%%%%%%%%%%%%%%%%%%%%%%%%%%%%%%%%%%%%%%%%%%%%%
%%%%% Figure: Mass vs Radius for EBs %%%%%
%%%%%%%%%%%%%%%%%%%%%%%%%%%%%%%%%%%%%%%%%%%%%%%%
%
%

%
%%%%%%%%%%%%%%%%%%%%%%%%%%%%%%%%%%%%%%%%%%%%%%%%
%%%%% EBs in SFRs %%%%%
%%%%%%%%%%%%%%%%%%%%%%%%%%%%%%%%%%%%%%%%%%%%%%%%
%
\section{Review of EBs in star-forming regions}
\label{book_EBs:EBs_SFRs}

The first low-mass EBs discovered in star-forming regions were identified in dedicated
long-term photometric surveys monitoring the Orion region \cite{stassun06, stassun07a, irwin07a},
one of the best studied area in the sky \cite{mjm96,hillenbrand97,hillenbrand00,feigelson03, hillenbrand13,ingraham14}. 
The first pair of eclipsing brown dwarfs (2MASS0535$-$05) at an age of about 1 Myr was reported 
by \cite{stassun06} with a period of about 10 days, an eccentric orbit, a significant mass ratio 
characterised by complementary high-resolution spectroscopy \cite{stassun07a}.
Above the hydrogen-burning limit, there two pairs of low-mass stars reported in the Orion Nebula 
Cluster.
%(Table \ref{tab_book_EBs:table_all_EBs}).
JW\,380 (=\,2MASS\,J05351214$-$0531388) 
was identified in the Monitor project 
\cite{irwin07a} with masses of 0.26 and 0.15 M$_{\odot}$ and in a period of 5.3 days. Par\,1802 
(=\,2MASS\,J05351114$-$0536512) was identified independently by several team. It is a pair of M4
twins with a mass of 0.4 M$_{\odot}$, a period just under 5 days and non-zero eccentricity. 
Despite similar masses, both components 
exhibit distinct temperatures and luminosities, suggesting that newborn binaries may differ 
in the physical properties as a result of their formation \cite{stassun08a,cargile08,gomez_maqueo12}.
In the ONC, we should add ISOY\,J0535$-$0447 announced by \cite{morales_calderon12} whose
masses are estimated rather than measured because they fixed the semi-major axis. The
primary is a K0 dwarf with a mass of 0.83 M$_{\odot}$ and a temperature of 5150\,K while
the secondary is most likely sub-stellar (0.05 M$_{\odot}$) but none have independent radius
measurements because the lines are not resolved spectroscopically. 
%Therefore, we list the system in Table \ref{tab_book_EBs:table_all_EBs} but it does not appear in the mass-radius plots.
These systems are key to test predictions from the theoretical pre-main-sequence models.

% In the ONC there is also ISOY J0535-0447 by Morales-Calderon et al. (2012): 0.83+0.05 Msun

%Low-mass binaries in Orion: JW380 and 2MASS0535$-$05
%\cite{irwin07a}, \cite{stassun06}, \cite{stassun07a} and \cite{stassun07b}
%
%Binaries in ONC from \cite{cargile08} and \cite{stassun08a}: Par\,1802, see also  \cite{gomez_maqueo12}

The advent of the K2 mission after the loss of one gyroscope of the Kepler satellite \cite{borucki10} 
led to the discovery of a handful of EBs over a wide range of masses in the nearest OB 
association to the Sun, Upper Scorpius (USco). %see table \ref{tab_book_EBs:table_all_EBs}. 
USco is located at 145 pc from the Sun and its age is currently debated in the literature, ranging between 
5 Myr and 10 Myr \cite{preibisch99,preibisch01,slesnick08,pecaut12,song12,pecaut16a,rizzuto16,david19a}
and subject to numerous photometric and spectroscopic surveys \cite{slesnick06,lodieu13c,dawson13}.
The first one in the M dwarf regime, UScoCTIO\,5 (2MASS\,J15595051$-$1944374), was selected 
as a photometric member by \cite{ardila00} later resolved as a spectroscopic binary by \cite{reiners05a}, 
and fully characterised by \cite{kraus15a} 
combining light curve from the photometry and high-resolution spectroscopy. A couple of other
low-mass M dwarf EBs with EPIC numbers have been identified in the K2 light curves
\cite{lodieu15c,david16a,david19a} as well as the first brown dwarf (RIK72\,=\,2MASS\,J16033922$-$1851297) 
orbiting a low-mass dwarf \cite{david19a}. A few other higher mass EBs have been reported 
at the age of USco but are not included in this review because we focus on the lowest mass objects \cite{alonso15a,david16a,david19a}. Nonetheless, we should emphasise that the EB
sequence of USco is fairly well constrained from high-mass stars down to the sub-stellar regime
thanks to the exquisite light curve delivered by K2 \cite{david19a}.
 
%At a slightly older age
%Low-mass and sub-stellar EBs in star-forming regions and associations such as Upper Scorpius.
%{\bf{NL}} \cite{lodieu15c}, 
%\cite{david16a} and \cite{alonso15a}.

At very young ages, we should also mention the discovery of the low-mass, pre-main sequence eclipsing 
binary, CoRoT\,223992193 (=\,2MASS\,J06414422$+$0925024), whose secondary lies at the K/M border
with a mass just under 0.5 M$_{\odot}$ and K dwarf primary with 0.67 M$_{\odot}$. This object belongs 
to the 3-6 Myr-old star-forming region NGC\,2264 that was monitored continuously for 23.4 days by the
CoRoT mission.

% Additional EBs in SFRs and clusters but more massive than 0.6 Msun
%    MML\,53               & $0.994\pm0.030$       & $0.857\pm0.026$       & \multicolumn{2}{c}{$2.201\pm0.071$\,$^b$}   & UCL         & 15        & 2010   & 12,13 \\ %[-0.7ex] Hebb2010a,Hebb2011
%    HD144548              & $0.984\pm0.007$       & $0.944\pm0.017$       & $1.319\pm0.010$     & $1.330\pm0.010$       & Upper Sco   & 5--10     & 2015   & 14 \\ %[-0.7ex]
%        V1174\,Ori            & $1.006\pm0.013$       & $0.7271\pm0.0096$     & $1.338\pm0.011$     & $1.063\pm0.011$       & Ori OB 1c   & 5--10     & 2004   & 15 \\  %[-0.7ex] Stassun2004
%    V818 Tau              & $1.06\pm0.01$         & $0.90\pm0.02$         & $0.76\pm0.01$       & $0.77\pm0.01$         & Hyades      & 600--800  & 2002   & 16 \\ %[-0.7ex]
%    RXJ\,0529.4$+$0041A   & $1.27\pm0.01$         &  $0.93\pm0.01$        & $1.44\pm0.10$       & $1.35\pm0.10$         & Ori OB 1a   & 7--13     & 2000   & 17,18,19 \\ %[-0.7ex]
%    NP Per                & $1.3207\pm0.0087$     & $1.0456\pm0.0046$     & $1.372\pm0.013$     & $1.229\pm0.013$       & Per OB 2    & 6--15     & 2016   & 20 \\ %[-0.7ex]
%    ASAS\,J0528$+$03      & $1.375\pm0.028$       &  $1.329\pm0.020$      & $1.83\pm0.07$       & $1.73\pm0.07$         & Ori OB 1a   & 7--13     & 2008   & 21 \\ %[-0.7ex]
% ISOY J0535-0447 & primary is a K0 with Teff = 5150K

%
%%%%%%%%%%%%%%%%%%%%%%%%%%%%%%%%%%%%%%%%%%%%%%%%
%%%%% EBs in open clusters %%%%%
%%%%%%%%%%%%%%%%%%%%%%%%%%%%%%%%%%%%%%%%%%%%%%%%
%
\section{Review of EBs in open clusters}
\label{book_EBs:EBs_OCs}

We can divide the stellar clusters targeted by K2 into two groups: the intermediate-age clusters
(100--200 Myr) whose main reference is the Pleiades with an age of 125$\pm$10 Myr 
\cite{stauffer98,barrado04b,mazzei89,dahm15,gossage18,lodieu19b} 
and the older more evolved open clusters like the Hyades (625$\pm$50 Myr) \cite{maeder81,mermilliod81,mazzei88,lebreton01,deGennaro09,brandt15a,martin18a,lodieu18b} 
and Praesepe (590--900 Myr) 
\cite{vandenberg84,mermilliod81,delorme11,gossage18,salaris04,bonatto04a,brandt15a,lodieu19b}.
All these clusters are within 200 pc \cite{babusiaux18} and have metallicities close to solar or slightly 
super-solar \cite{boesgaard90,cayreldestrobel97,grenon00}

\begin{figure}
\sidecaption
% For example, with the graphicx style use
\includegraphics[width=\linewidth]{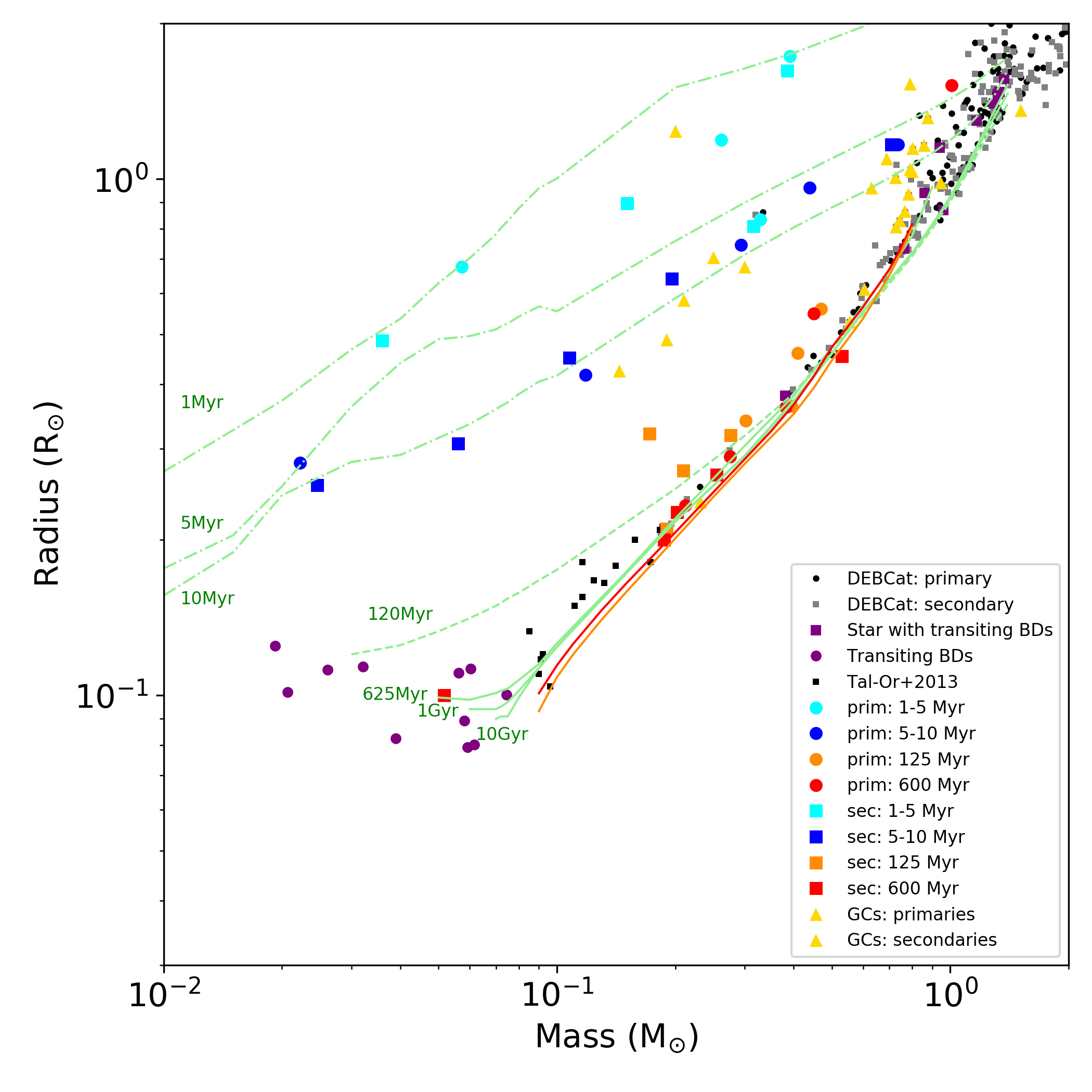}
%
% If no graphics program available, insert a blank space i.e. use
%\picplace{5cm}{2cm} % Give the correct figure height and width in cm
%
\caption{Mass-radius relation for low-mass and sub-stellar eclipsing binaries in
star-forming regions, open clusters, and globular clusters.
The primaries and secondaries are plotted as dots and squares, respectively.
Colour scheme as follows: 1--5 Myr (cyan), 5--10 Myr (blue); 125 Myr (orange); 600 Myr (red);
globular clusters (yellow).
The primaries and secondaries of EBs from the DEBcat database whose accuracies are 
better than 2\% on their masses and radii are displayed as black and grey symbols.
Overplotted with green lines are the BT-Settl isochrones for ages of 1 Myr, 5 Myr, 10 Myr, 
120 Myr, 625 Myr, 1 Gyr, and 10 Gyr. We added the 5 Gyr-old low-metallicity tracks at
[M/H]\,=\,$-$2.0 and $-$1.0 dex as orange and red lines, respectively.}
\label{fig_book_EBs:fig_M_R_EBs}
\end{figure}

In the Pleiades, HII\,2407 known as a single-lined EB was identified eclipsing every 7.05 days 
\cite{david15b}. The primary is well characterised with a spectral type of K1-K3, an effective
temperature of 4970$\pm$95\,K, a mass of 0.81$\pm$0.08 M$_{\odot}$, and a radius of 
0.77$\pm$0.13 R$_{\odot}$. The secondary is a low-mass M dwarf, undetected in high-resolution 
spectra shortwards of 800 nm, yielding mass and radius estimates of 0.18 M$_{\odot}$ and 
0.21 R$_{\odot}$, respectively. Another two EBs members of the Pleiades have been
characterised photometrically and spectroscopically: HCG\,76 and MHO\,9 \cite{david16c}.
Both systems have long orbital periods with masses and radii well determined from the K2 light curve
and multiple radial velocity epochs. Two other higher mass EBs are presented in
\cite{david16c} as well as a possible member but not discussed in this review focusing on low-mass
dwarfs. Finally, we highlight the possibility of MHO\,9 being a hierarchical triple system due to its
position above the Pleiades sequence in the H-R diagram (Fig.\ \ref{fig_book_EBs:fig_M_R_EBs}).

A pair of M dwarfs (2MASS\,J04463285$+$1901432)  with a short period ($\sim$0.62 days) was reported 
by \cite{hebb06} with masses of 0.47$\pm$0.05 and 0.19$\pm$0.02 M$_{\odot}$ in the 150 Myr-old cluster 
NGC\,1647 \cite{dias02a} located at 540 pc from the Sun \cite{turner92a}. The system is confirmed as a 
photometric and spectroscopic member with a radial velocity consistent with the mean value of the cluster.
This new low-mass system represent an important link between the Pleiades and older clusters 
discussed below.

% Pleiades: \cite{david15b}, \cite{david16c}

% In NGC\,1647 at 150 Myr an EB by \cite{hebb06}.

Four low-mass EBs have been revealed in the Praesepe cluster. PTFEB132.707$+$19.810
was announced by \cite{kraus17a} as a pair of 0.38$+$0.20 M$_{\odot}$ going around each 
other every 6 days and independently announced by \cite{gillen17a} as AD\,3814\@.
Another three EBs were included in the sample of low-mass EBs discussed in \cite{gillen17a}. 
Two of these cluster candidates were classified as Praesepe members by four of six surveys \cite{adams02a,kraus07d,baker10,boudreault12,khalaj13,wang14a}, while the fourth one (AD\,1508) is only labelled as member in two of these surveys. AD\,2615 is a pair of 
almost equal-mass M dwarfs (0.21$+$0.25 M$_{\odot}$) with a period of 11.6 days and no eccentricity. The most special system, AD\,3116, is composed
of a M dwarf ($\sim$0.28 M$_{\odot}$) and a brown dwarf with an estimated mass of 0.052 M$_{\odot}$. The period of this system is quite short, around 2 days, and this is the only system with an
significant eccentricity of 0.142\@. The last system, with a doubtful membership, is composed of two low-mass dwarfs close to the 0.5 M$_{\odot}$, limit set in this review, revolving every 1.55 days. For a complete census of eclipsing systems in the Beehive cluster, we should highlight the seven transiting exoplanets, including three orbiting members with masses equal or below 0.5 M$_{\odot}$
\cite{mann17a}. These four system are unambiguously confirmed as astrometric members of
the Praesepe cluster from the 3D kinematic selection using the second release of $Gaia$ \cite{lodieu19b}.

% Praesepe 1 in PTF \cite{kraus17a}: PTFEB132.707+19.810
% 4 EBs in Praesepe from \cite{gillen17a}

% Hyades \cite{david16c} + \cite{mann16a}

In the Hyades, no low-mass EB was disclosed in the K2 light curves. However, one transiting
system member of the Hyades \cite{vanAltena66a,hanson75} was announced independently
by \cite{david16c} and \cite{mann16a}. The primary, vA\,50, has a Neptune-size planet with 
an upper limit on its mass of 1.1 Jupiter mass based on high-resolution spectroscopic radial 
velocity with an accuracy of around 0.3 km/s. This planet is orbiting a M dwarf member of the 
Hyades with a mass of 0.26 M$_{\odot}$ and a radius of 0.32 R$_{\odot}$ every $\sim$3.5 days. 
This object is confirmed as a $Gaia$ astrometric member located at 4.63 pc from the center
of the Hyades cluster in 3D space \cite{lodieu19a}.

%
%%%%%%%%%%%%%%%%%%%%%%%%%%%%%%%%%%%%%%%%%%%%%%%%
%%%%% Figure: Mass vs Radius for EBs %%%%%
%%%%%%%%%%%%%%%%%%%%%%%%%%%%%%%%%%%%%%%%%%%%%%%%
%
%

%
%%%%%%%%%%%%%%%%%%%%%%%%%%%%%%%%%%%%%%%%%%%%%%%%
%%%%% EBs in globular clusters %%%%%
%%%%%%%%%%%%%%%%%%%%%%%%%%%%%%%%%%%%%%%%%%%%%%%%
%
\section{Review of EBs in globular clusters}
\label{book_EBs:EBs_GCs}

The globular clusters of our Milky Way are the oldest objects we know of.
They are very massive containing up to a million of individual stars. There are
two distinct populations of globular clusters, namely the classical population in the Galactic 
Halo \cite{Baumgardt2018} and the one in the Galactic Bulge \cite{Rossi2015}. The latter is 
younger (ages about 10 Gyr) and more metal-rich ($-$0.7$<$\,[Fe/H]\,$<$\,$+$0.5 dex) than the one 
in the Galactic Bulge \cite{Bica2006}. The H-R diagram for a typical globular cluster looks very 
different than that of an open cluster. There are no main-sequence stars of spectral types earlier 
than F, but there are many red giants and other objects of the late evolutionary phase 
(for example the horizontal branch) of low-mass stars. 

Since the first discovery of different internal populations (of the main-sequence as well as the giant 
branches) of stars in globular clusters \cite{Piotto2007} using the Hubble Space Telescope, this characteristic
was found for almost all known aggregates. However, there are no unique pattern or correlation with other 
astrophysical parameters known so far. The only possible explanation of the observations is a different 
(enriched) helium abundance of these internal populations \cite{Cassisi2017}.
However, the evolutionary mechanism behind this enrichment of helium is still unknown. 

From an observational point of view, globular clusters are difficult to observe because they are very dense 
(up to 100 stars per arcsec$^{2}$), typically far away (several kpc) from the Sun, and the low-mass members 
have apparent magnitudes fainter than 20$^{th}$ magnitude. In order to resolve most of the cluster areas, 
large ground-based telescope and good seeing conditions or satellite measurements are needed. Especially 
time-series of radial velocity measurements are almost not available. 

In a series of papers, the Clusters Ages Experiment (CASE; \cite{Kaluzny2007,Kaluzny2015}) investigated 
photometrically and spectroscopically several eclipsing binary systems in different globular clusters. 
Most of these systems are so-called blue stragglers which are more luminous and bluer than stars at the 
main-sequence turnoff point for their host cluster \cite{Sandage1953}. Therefore, these objects
are the brightest main-sequence stars in the cluster and easier to observe. However, these eclipsing binary 
systems are peculiar in the sense that normally a significant interaction between the components took place. 
For example, there is a scenario in which the primary component is reborn from a former white dwarf that 
accreted a new envelope through mass transfer from its companion. The secondary star has lost most of its 
envelope while starting its ascent onto the sub-giant branch. It failed to ignite helium in its core and is 
currently powered by an hydrogen-burning shell \cite{Kaluzny2007}. The time scales of the different stages 
of all these processes are not known. Analysing the individual components in the mass versus radius diagram 
(Fig.\ \ref{fig_book_EBs:fig_LC_RV_EBs}) might help putting further constraints on the models.

The estimation of the binary fraction of globular cluster members is severely influenced by the above 
described observational constraints. \cite{Sollima2007} investigated the fraction of binary systems in 
a sample of 13 low-density Galactic globular clusters using Hubble Space Telescope observations. They 
analysed the colour distribution of main-sequence stars to derive the minimum fraction of binary systems 
required to reproduce the observed colour-magnitude diagram morphologies. They found that all the analysed 
globular clusters contain a minimum binary fraction larger than 6\% within the core radius. However, 
the estimated global fractions of binary systems range from 10 to 50\% depending on the cluster. 
More recently, \cite{Ji2013} determined the binary fractions for 35 globular clusters using different 
models including a star superposition effect. They derived a binary fraction of 6.8\% to 10.8\% depending 
on the assumed shape to the binary mass-ratio distribution, with the best fit occurring for a binary 
distribution that favours low mass ratios (and higher binary fractions). Later on, \cite{Lucatello2015} 
presented a long-time observational campaign using FLAMES spectra of 968 red giant branch stars located 
around the half-light radii in a sample of ten Galactic globular clusters. From these only 21 radial 
velocity variables were identified as bona-fide binary stars, yielding a binary fraction of 2.2\,$\pm$\,0.5\%. 
Finally, \cite{Milone2016} found a binary fraction between 3\% and 38\% depending on the regions of eight 
globular clusters. This short overview shows already the wide range of derived values and the need for 
a new homogeneous analysis of all available photometric and spectroscopic data.

The newest version (November 2017) of the variable stars in Galactic globular clusters catalogue 
\cite{Clement2001} was used to estimate the percentage of eclipsing binary systems in respect to all known 
variables. Here, we want to recall that in these old aggregates, we find mainly pulsating variables, 
such as Cepheids, Giants, SX Phoenicis, RR Lyrae, and RV Tauri stars. The pulsational characteristics 
(i.e.\ periods as well as amplitudes) and astrophysical driving mechanism are widely different \cite{Percy2011}. 
Nevertheless, the amplitudes of these stars are comparable to those of eclipsing binaries which should not 
introduce a significant bias in the detection rate. The mentioned catalogue includes 5604 stars in 151 globular
clusters. In total, 399 eclipsing binaries of all types (excluding field stars) are listed. The distribution 
of the apparent magnitudes ranges from 12 to 24 with a peak at 17.5 mag, respectively. To put this number 
in a broader context, we need an estimate of the total number of investigated stars per cluster and thus 
the overall variability ratio. This number crucially depends on the telescope used, time series characteristics,
and the methods applied to analyse time series. To get a rough estimate of this number, we use five recent 
publications. In the following, we list the total number of observed stars, the included variable stars 
(known, new, and suspected), and the eclipsing binaries: 7630/40/1 \cite{Deras2019}; 4274/59/0 \cite{Yepez2018}; 
132457/359/30 \cite{Rozyczka2017}, 31762/47/1 \cite{Tsapras2017}; and 11358/13/1 \cite{Lee2016}.
The number of detected variables in globular clusters is only a few percent from which only a maximum of 
10\% are eclipsing binaries. Therefore, also in the future the number of known eclipsing binary systems will
not significantly increase. To identify possible eclipsing binary systems with low-mass companions, 
available light curves have to be analysed and the best candidates for spectroscopic follow-up 
observations selected.

%
%%%%%%%%%%%%%%%%%%%%%%%%%%%%%%%%%%%%%%%%%%%%%%%%
%%%%% Discussion %%%%%
%%%%%%%%%%%%%%%%%%%%%%%%%%%%%%%%%%%%%%%%%%%%%%%%
%
\section{Discussion}
\label{book_EBs:discussion}
%

%
%%%%% Frequency %%%%%
%
\subsection{Frequency of EBs}

Most of the low-mass EBs identified so far in star-forming regions and open clusters come from
the CoRoT and Kepler space missions, except for those
members of Orion. Here in this section we provide a tentative estimate of the frequency of low-mass
EBs in the regions investigated so far to spot any potential trend with age or environment.  

Several studies looked at the fraction of spectroscopic binaries in the M dwarf regime. The first study
f detected variables in globular clusters is only a few percent from which only a maximum of
10\% are eclipsing binaries. Therefore, also in the future the number of known eclipsing binary systems will
not significantly increase. To identify possible eclipsing binary systems with low-mass companions,
available light curves have to be analysed and the best candidates for spectroscopic follow-up
observations selected.

of M dwarf multiples revealed about 1.8$\pm$1.8\% of spectroscopic binaries (0.04$-$0.4 au) for a 
small sample of a few tens of low-mass stars \cite{fischer92}. The CARMENES team identified nine 
double-line spectroscopic binaries with periods in the 1.13$-$8000 days interval among their sample of 
342 M dwarfs, yielding a multiplicity of 2.6\% \cite{baroch18}. The search for spectroscopic binaries in 
the Sloan database returned 3$-$4\% of multiple systems with separations less than 0.4 au with a 
possible towards the hottest M dwarfs \cite{clark12a}. At later spectral types, the frequency of 
spectroscopic binaries among late-M dwarf (M5$-$M8) binaries is around 11\% for separations
in the 0$-$6 au range \cite{basri06}, while an independent survey of 58 M8$-$L6 dwarfs yielded a 
0.9$-$11.1\% multiplicity at separations closer than 1 au \cite{blake10a}. Lastly, we should mention
the statistical occurrence of M dwarf systems in the Kepler field of view of 7$-$13\% based on the
fractional incidence of low-mass eclipsing binaries \cite{shan15a}.

\begin{figure}
%\sidecaption
\centering
% For example, with the graphicx style use
%\includegraphics[width=0.48\linewidth]{plot_clusters_EBs_Mass_a.png}
%\includegraphics[width=0.48\linewidth]{plot_clusters_EBs_Mass_Ecc.png}
\includegraphics[width=0.85\linewidth]{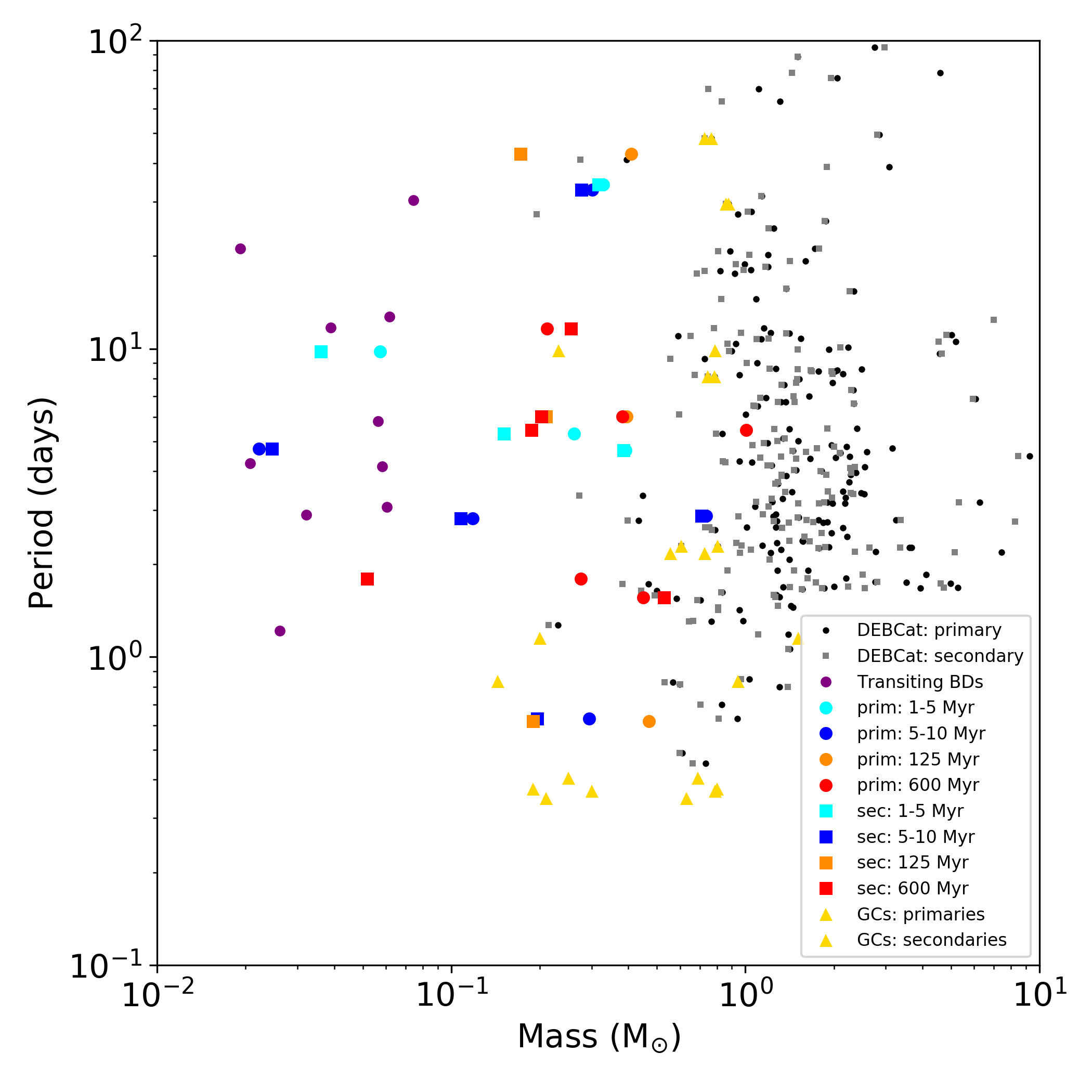}
\includegraphics[width=0.85\linewidth]{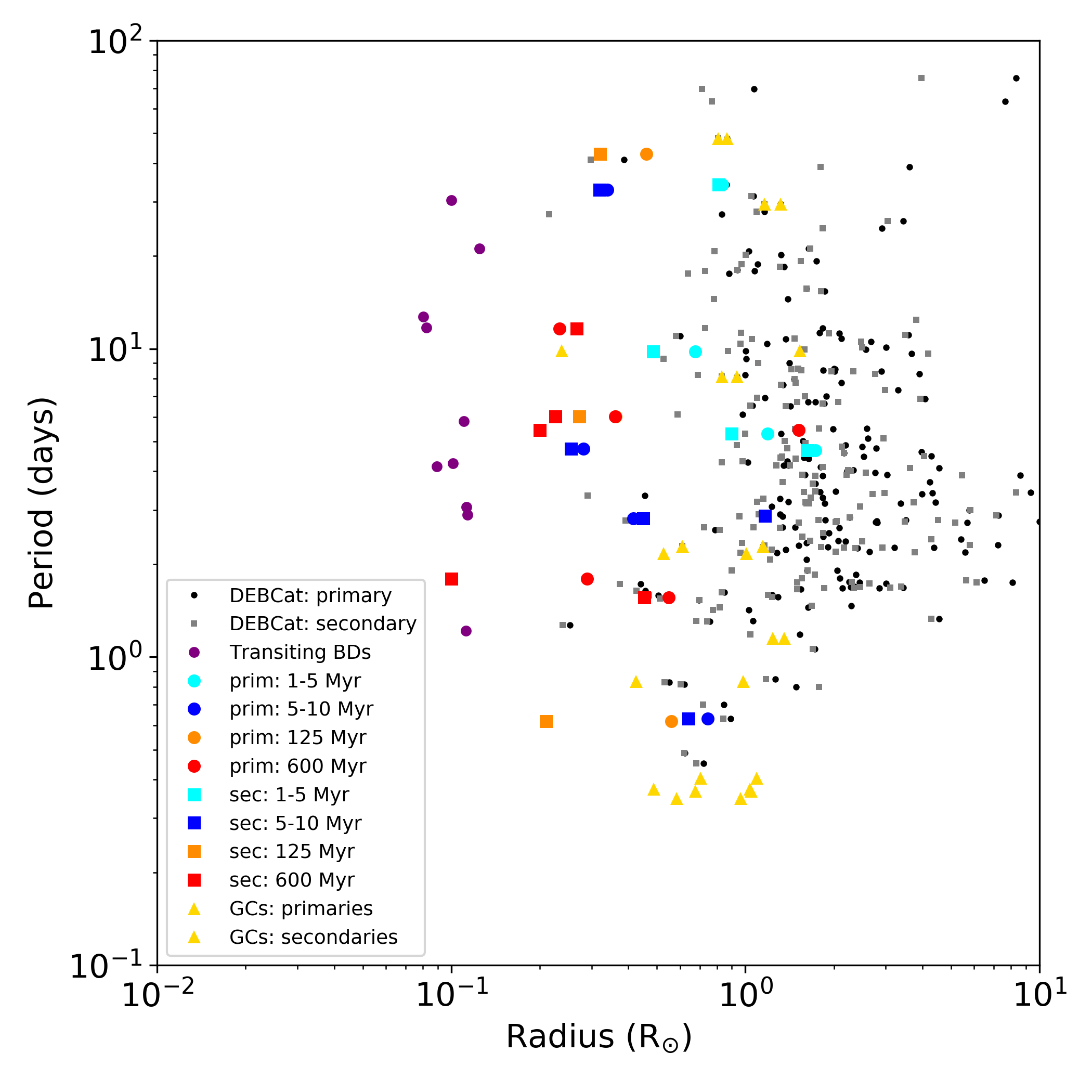}
%
% If no graphics program available, insert a blank space i.e. use
%\picplace{5cm}{2cm} % Give the correct figure height and width in cm
%
\caption{The mass-period and radius-period diagrams of low-mass EBs: 
The primaries and secondaries are plotted as dots and squares, respectively.
Colour scheme as follows: 1--5 Myr (cyan), 5--10 Myr (blue); 125 Myr (orange); 
600 Myr (red), globular clusters (yellow), field EBs from the DEBcat database (black$+$grey).}
\label{fig_book_EBs:M_R_Period}
\end{figure}

While the Kepler mission monitored a single field towards the Cygnus constellation, the K2 mission
targeted star-forming regions and open clusters in the ecliptic for periods of approximately consecutive 
80 days, corresponding to semi-major axis less than a\,=\,0.36 au. However, if we assume that a minimum
of two transits are necessary to identify any eclipsing binaries with high confidence, searches in K2 would 
be sensitive to periods less than about 50 days, i.e.\ a\,$\leq$\,0.25 au (Fig.\ \ref{fig_book_EBs:M_R_Period}). 

\begin{figure}
%\sidecaption
\centering
\includegraphics[width=0.85\linewidth]{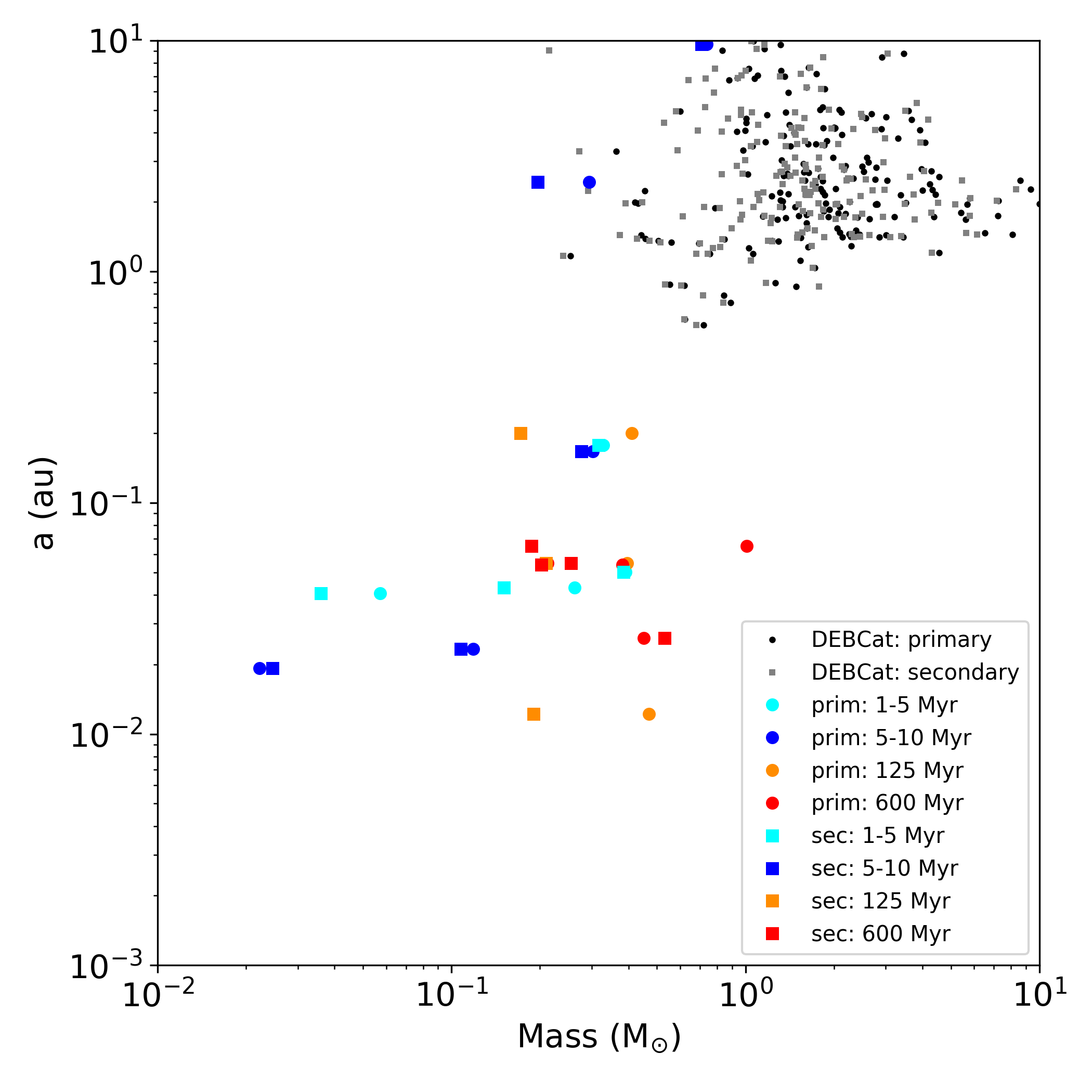}
\includegraphics[width=0.85\linewidth]{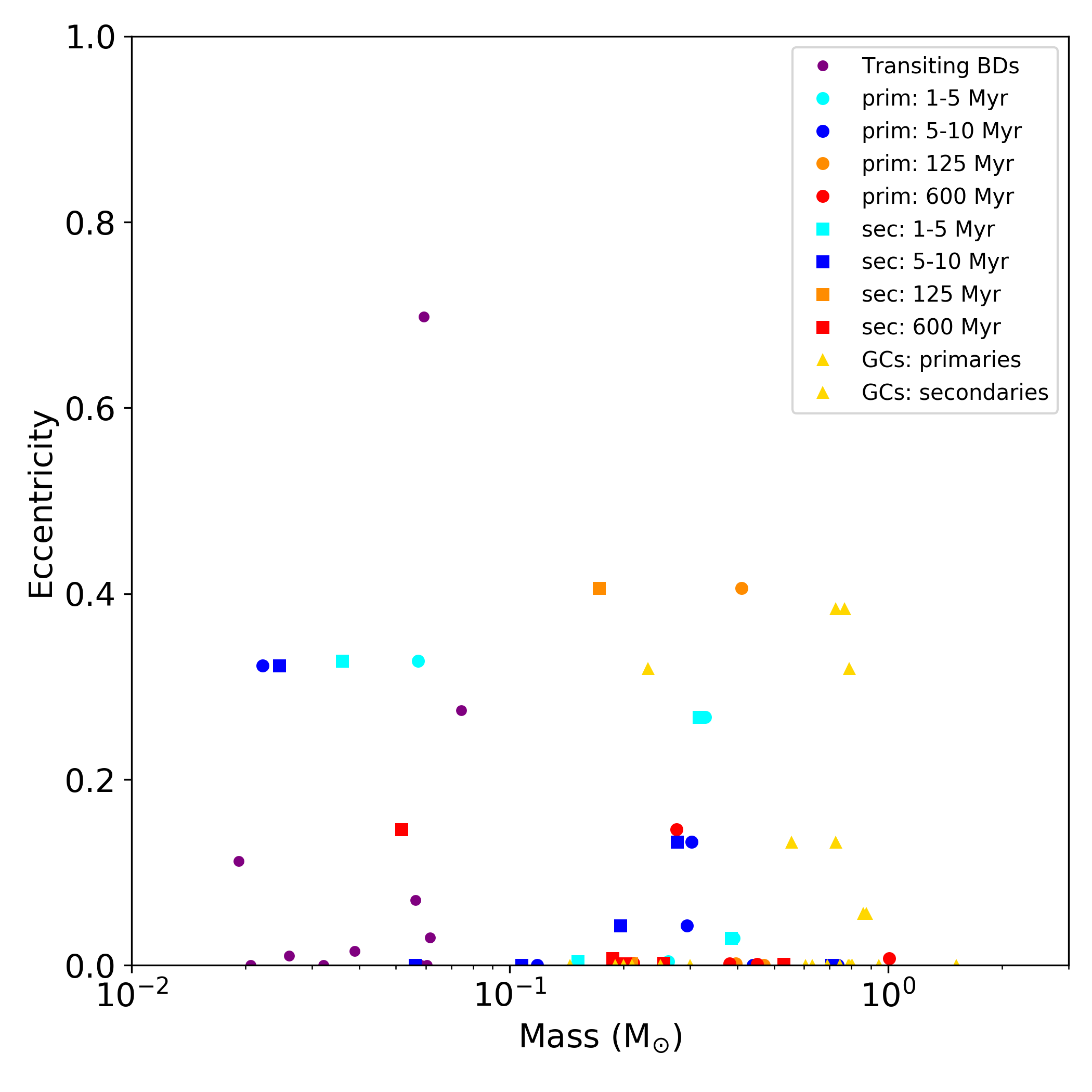}
\caption{The mass-semi-major axis and mass-eccentricity diagrams of low-mass EBs
in star-forming regions and clusters. 
The primaries and secondaries are plotted as dots and squares, respectively.
Colour scheme as follows: 1--5 Myr (cyan), 5--10 Myr (blue); 125 Myr (orange); 
600 Myr (red), globular clusters (yellow), field EBs from the DEBcat database (black$+$grey).}
\label{fig_book_EBs:M_a_e}
\end{figure}

The census of low-mass EBs in Upper Scorpius, the Pleiades, Praesepe, NGC\,2264, and Ruprecht\,147
is 6, 2, 5, 1, and 2, respectively. These numbers represent lower limits for several reasons intrinsic to
the search for eclipsing systems: incompleteness of the samples, inhomogeneous quality of the light-curves
depending on the brightness of the targets, lack of sensitivity to large mass ratios, etc\ldots{}.
The rotation properties of M dwarf members of Upper Scorpius, the Pleiades, and Praesepe have 
been investigated in great details \cite{rebull16a,stauffer16a,rebull17,rebull18} thanks to K2\@.
Using a crude selection of potential M dwarfs with effective temperature below 3800\,K and $V-K_{s}$
colours redder than 3.8 mag, we identified 867, 566, 619 low-mass members in Upper Scorpius, 
the Pleiades, Praesepe, respectively. We derived a frequency of EBs with semi-major axis less than
$\sim$0.25 au of 0.64\%, 0.35\%, and 0.8\% in these three regions. We estimate uncertainties up to
50\% because of the low number statistics of published EBs, the rough photometric selection of M
dwarf members, and the level of contamination of ground-based surveys before the advent of $Gaia$.
Overall, we can argue that the frequency of EBs in clusters is below 1\% for separations less than 0.25 au
with no significant variation with age or environment. We also show the mass-semi-major axis and
mass-eccentricity diagrams for cluster EBs in Fig.\ \ref{fig_book_EBs:M_a_e}. \\
%{\bf{Need to figure out the numbers of M dwarfs in NGC\,2264 and Ruprecht\,147}}.

Finally, we should mention that only one brown dwarf pair is known eclipsing \cite{stassun06,stassun07b},
making any estimate of the frequency of sub-stellar EBs in young regions unreliable statistically. However, 
these types of systems should be rare although we cannot discard observational biases due to their 
intrinsic faintness and the lack of long-term monitoring sensitive to the sub-stellar population in star-forming 
regions and open clusters.

%
%%%%%%%%%%%%%%%%%%%%%%%%%%%%%%%%%%%%%%%%%%%%%%%%
%%%%% Figure: Diagrams with Periods %%%%%
%%%%%%%%%%%%%%%%%%%%%%%%%%%%%%%%%%%%%%%%%%%%%%%%
%
%

%
%%%%%%%%%%%%%%%%%%%%%%%%%%%%%%%%%%%%%%%%%%%%%%%%
%%%%% Figure: Mass vs a && Mass vs Ecc %%%%%
%%%%%%%%%%%%%%%%%%%%%%%%%%%%%%%%%%%%%%%%%%%%%%%%
%
%

%
%%%%% Age %%%%%
%
\subsection{The impact of age on mass and radius}
\label{book_EBs:age}

Fig.\ \ref{fig_book_EBs:fig_M_R_EBs} clearly shows that age has a strong impact on the
radius of M dwarfs younger than $\sim$500 Myr: the younger the M dwarf, the larger is its 
radius, as predicted by evolutionary models \cite{burrows93,baraffe98,baraffe15}.
At a given mass, the radius of a Pleiades M dwarf at 120 Myr is about 10\% larger than
the radius of a Praesepe or a Hyades member (600$-$700 Myr), which is comparable to the ones
of older field stars (ages\,$>$\,1 Gyr). We do not see any difference at ages older than 500 Myr 
for low-mass M dwarfs in the H-R diagram. Models do predict differences in the sub-stellar regime 
but only one brown dwarf with an age larger than 500 Myr has been reported in Praesepe.
The difference is small going from 600 Myr to 120 Myr but significant moving towards much
younger ages: the radius of a 0.25 M$_{\odot}$ is approximately 3 and 5 times larger at
5$-$10 and $\sim$3 Myr, respectively. We observe a clear difference in radii of M dwarf
members of Upper Scorpius (5$-$10 Myr; blue symbols) with those in Orion ($<$3 Myr; cyan) 
compared to those of the Pleiades (125 Myr; orange symbols). We note that one EB system 
identified in NGC2264 \cite{gillen14} confirms that the age of the cluster lies between the age 
of Orion and Upper Scorpius based on its location in the H-R diagram displayed in 
Fig.\ \ref{fig_book_EBs:fig_M_R_EBs}. We also remark that the system found in NGC\,1647
\cite{hebb06} whose age is constrained to 150$\pm$10 Myr lies slightly above the evolutionary
model at 120 Myr (Pleiades-like age), suggesting that a revision of the age of NGC\,1647
(and possibly its distance checking the parallaxes of $Gaia$ DR2) might be needed.

We also investigated the dispersion of the 11 eclipsing brown dwarfs (purple dots in
Fig.\ \ref{fig_book_EBs:fig_M_R_EBs}) revealed by several missions and ground-based surveys.
This dispersion might be the consequence of tides from the primary star yielding engulfment of
the companions, magnetic activity, presence of cold spots on the surface of the brown dwarf, 
irradiation from the host star, and/or metallicity. The puzzle remains, however, under debate. 
Because of the improved knowledge on the impact of the age on the radius of M dwarfs and
brown dwarfs gathered over the past years mainly thanks to Kepler K2, we collected information
on the ages of the primary stars from the discovery papers to compare the values with the
inferred from the latest BT-Settl isochrones \cite{baraffe15}. We considered
the effect of age in this review keeping in mind current uncertainties on the mass determinations
of the sub-stellar companions arising from the uncertainties on the mass of the primary and
the grazing transits of some of the examples. 

First of all, we note that two of these brown dwarfs orbit a M dwarf primary. Based on the above 
discussion, we can argue that the ages of these M dwarfs are older than 100 Myr although any 
older age is possible when taking into account the uncertainties on their masses and radii. 
Comparing the positions of the eclipsing brown dwarfs to the latest BT-Settl isochrones, we divided 
the sample into several groups. 
Two systems (KOI\,415 and LHS\,6343) appear very old, older than the others due to their small 
radii, consistent with the analysis of the discovery papers. We note again the low metallicity
of KOI\,415 ([Fe/H]\,=\,$-$0.24 dex) suggesting an old age.
Another three systems (WASP\,30, KOI\,189, KOI\,205) appear old, with ages around 1 Gyr or older
\cite{anderson11a,diaz14a,bonomo15}.
The positions of CoRoT-15b and CoRoT-33b in the H-R diagram fit well the 300 Myr-old isochrone.
However, we caution this point because the masses of the secondaries are ill-defined due to
the intrinsic faintness of CoRoT-15 ($V$\,$\sim$\,16 mag) and the grazing eclipse of CoRoT-33b
\cite{bouchy11a,csizmadia15}. Nonetheless, those two systems appear as intermediate age-wise
between the aforementioned systems and the (possibly) youngest ones described below.
The last group of brown dwarfs exhibit inflated radii with respect to their siblings, the most extreme
one being Kepler-39b \cite{siverd12}. Its age remains controversial depending on the method used
for its determination: fit to the spectrum of the solar-type primary suggests 1.$-$2.9 Gyr while the gyrochronology age infers 0.4$-$1.6 Gyr \cite{mamajek08}. The brown dwarf is 
best fit by isochrones with ages between 50 and 120 Myr. The masses and radii of the brown dwarfs in 
the other systems (NLTT\,41135, KELT-1b, and CoRoT-3b) are best fit by isochrones with ages
bracketed by the Pleiades and Hyades isochrones \cite{irwin10,siverd12,deleuil08}. 
We emphasise that three of the solar-type primaries appear over-luminous compared to the 
others and the BT-Settl isochrones. Overall, in spite of the current uncertainties on the masses
and radii of the sub-stellar secondaries, we cannot discard age to have a significant effect on
their radii due to the dependence of physical properties of brown dwarfs with gravity
\cite{gorlova03,gagne14a,filippazzo15,martin17a,lodieu18a}. The revision of the distances
of the host stars with the $Gaia$ parallaxes should revise some of these discrepancies and
decrease current error bars.

%
%%%%% Metallicity %%%%%
%
\subsection{The impact of metallicity on mass and radius}
\label{book_EBs:FeH}

It is widely established that the fraction of stars hosting planets is larger with higher metallicity.
The average metallicity of a volume-limited sample of stars with planets that have been specifically 
searched for planets peaks at [Fe/H]\,$\sim$\,$+$0.1 \cite{santos05a,bond06}.
The frequency of metal-poor stars with planets is of the order of 5\% or less, while more
than 30\% of metal-rich stars ($\geq$\,$+$0.25 dex) host planets. Moreover, there might be a 
trend towards low-mass planets with short periods orbiting low metallicity stars
\cite{santos03a,pinotti05}.

We should compare the different frequencies and see whether metallicity has an impact
\cite{El_Badry19a}.

The spectra of M dwarfs start to be affected by the dearth of metals at optical and near-infrared 
wavelengths for metallicities below [Fe/H]\,$\sim$\,$-$0.5 dex \cite{gizis99,lepine07c,burgasser07b,jao08},
trend extending towards at temperatures below 2500\,K \cite{kirkpatrick16,zhang17a}.
The impact of metallicity on the sample of low-mass EBs in star-forming regions and open 
clusters is hard to disentangle from the effect of age because all regions have a metallicity 
equal or close to solar within 0.2 dex. 

We also looked at the possible impact of metallicity on the dispersion of eclipsing brown dwarfs in the 
H-R diagram (Fig.\ \ref{fig_book_EBs:fig_M_R_EBs}). Among this sample, only two stars
stand out due to their metallicity: CoRoT-33 a G9V with [Fe/H]\,=\,$+$0.44$\pm$0.10 dex \cite{csizmadia15} 
and KOI\,415 a G0IV metal-poor solar-type analogue with [Fe/H]\,=\,$-$0.24$\pm$0.10 dex \cite{moutou13}.
The latter is not so different from the bulk of stars with eclipsing sub-stellar companions because
the difference in metallicity is less 0.2 dex but we note that its orbit is eccentric and its
radius among the two lowest. CoRoT-33 is classified as a old star with an age greater
than 4.6 Gyr based on a serie of indicators \cite{csizmadia15}. The radius of the brown dwarf is
40\% larger than KOI\,415 for an almost identical mass (59 M$_{\rm Jup}$ vs 62 M$_{\rm Jup}$). 
Based on this comparison, we conclude that metallicity may indeed play a role in the properties of sub-stellar
objects, in line with the spectral differences seen in L and T subdwarfs \cite{kirkpatrick16,zhang17a}.

%
%%%%% Stellar activity %%%%%
%
\subsection{The role of stellar activity and activity cycles}
\label{book_EBs:activity}

The 11 year-long activity cycle on the Sun is known for a
long time. The first long-term brightness changes which were
interpreted as starspot cycles for M-type stars
were reported by \cite{Phillips1978}. Chromospheric activity of F- to M-type 
stars can be studied using long-term Ca\,H\&K data, for example
from the Mount Wilson survey \cite{Wilson1978}. Those observations show 
cyclic variations yielding relations between the rotational period, the length
of the activity cycle, and other stellar properties. Most important,
faster rotating stars have shorter activity cycles \cite{Baliunas1996},
which can be explained by the classical dynamo theory. The square of the ratio 
of the cycle length and the rotational period can be used as a quantity to parametrise
activity cycles.

In many active stars the starspots are so large that they cause brightness variations
which can be few tens of percent from the mean light level \cite{Lockwood2007}, thus 
making them easily observable. The observed cycle lengths seem to converge with stellar 
age from a maximum dispersion around the Pleiades' age towards the solar cycle value 
at the Sun's age \cite{Messina2002},
and that the overall short- and long-term photometric variability increases with inverse
Rossby number. The cycles of active stars are often not as regular and cyclic as 
their more quiet counterparts. Many active stars exhibit multiple cycle lengths simultaneously and cycle lengths in active stars are also often variable \cite{Olah2009}. Most intriguing is the so-called flip-flop phenomenon where the activity concentrates 
on two permanent active longitudes, and flips between the two every few years \cite{Jetsu1993}.

The phenomenon are also important for the interpretation of light curves for eclipsing binary systems. The chromospherically active components of 180 low-mass pre-main sequence stars and chromospherically active binary systems have been looked at by \cite{Parihar2009} and \cite{Eker2008}, respectively. The light curves of such systems are complicated to interpret if one or even two components show spots on different time 
scales and with different intensities. The effects of starspots on the light curves 
of eclipsing binaries, and, in particular, how they may affect the accurate measurement 
of eclipse timings have been investigated by \cite{Watson2004}. For systems containing 
a low-mass main-sequence star and a white dwarf, the times of primary eclipse ingress 
and egress can be altered by several seconds (larger effect for lower inclinations) for 
typical binary parameters and star-spot depressions. These effects cause a jitter in the residuals of O$-$C diagrams, which can also result in the false detection of spurious orbital period changes.

A nice example of how to model a light curve taking account of all the above mentioned effects is presented by \cite{Czesla2019}, who investigated the short-period (2.17 d) eclipsing binary CoRoT\,105895502\@. They found a starspot with a period of about 40 days which remains quasi-stationary in the binary frame, and one starspot showing prograde motion at a rate of 2.3 degree per day, whose lifetime exceeds the duration of the observation (145 days). Only with eclipsing binary systems it is possible to study the complex correlations between chromospheric activity, spot cycles, and the astrophysical parameters in more details.

\subsection{Flares in M dwarfs}
\label{book_EBs:activity_flares}
Flares are mostly rapid transients lasting of the order of minutes or dozens of minutes, which are observable in different regions of the electromagnetic spectrum, from the ultra-violet to the X-ray domains. They are often observed on M dwarfs because they ideally fulfill the necessary conditions. Firstly, they are low mass stars with magnetic fields that remain active for a substantial part of their lives, and, secondly, the large difference between hot flaring regions and the cool photosphere give a higher chance of magnetic field generation. The energy of flares on M dwarfs can reach 10$^{28}$--10$^{29}$ W. For example, on 2014 April 23, the SWIFT satellite detected a super-flare from the nearby young M-type binary DG\,CVn with the radiated energy about 4--9$\times$10$^{28}$ W of energy in the 0.3--10 keV X-ray bandpass. This is about 10,000 times stronger than the most powerful solar flare on record \cite{osten2016}. The current available large surveys like ASAS-SN, Next Generation Transient Survey (NGTS), Kepler/K2, and TESS provide excellent photometric data to study the occurrence and frequency of flares on M dwarfs (e.g.\ \cite{davenport2016}, \cite{Ilin2019}, \cite{vandoor2017}, \cite{hawley2014}, \cite{gunther2019}, \cite{jackman2019}, \cite{schmidt2019}). Furthermore, selected M dwarfs with observed flares were also monitored spectroscopically \cite{chang2017} to determine their ages and study the influence of the flare on the presence of possible exoplanets orbiting M dwarfs. 

\begin{figure}
\centering
\includegraphics[width=\linewidth]{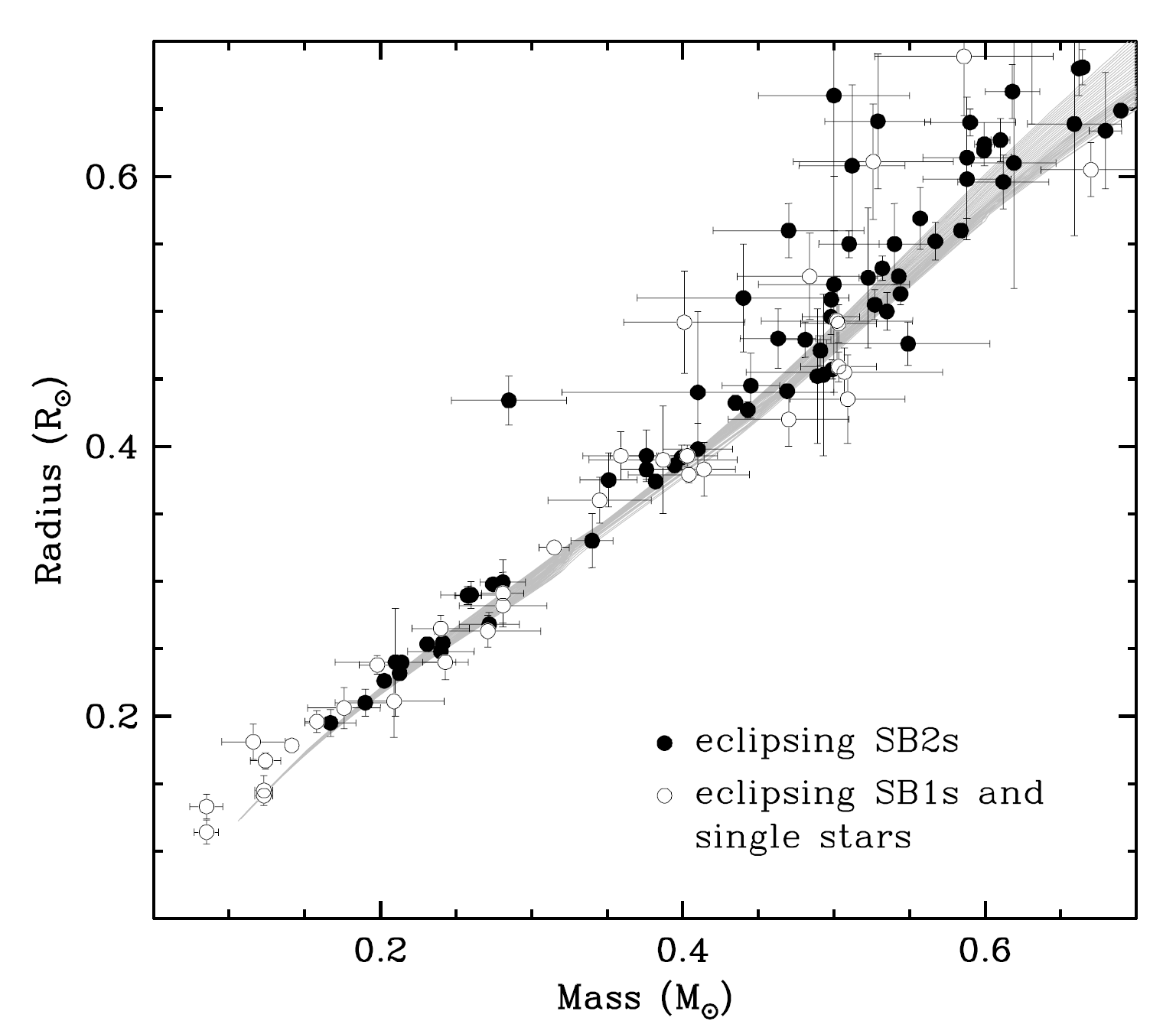}
\caption{Mass-radius diagram for low-mass stars, including
all measurements for double-lined eclipsing binaries (SB2;
filled symbols) as well as determinations for single-lined
eclipsing systems (SB1) and single stars (open symbols).
Solar-metallicity Dartmouth isochrones are shown for comparison,
for ages ranging from 1 to 13 Gyr (grey band) \cite{torres2013}.}
\label{fig_book_EBs:mass_rad}
\end{figure}

It was proposed that flares together with dark magnetic spots are responsible for the difference between the observed radii of M dwarfs and predicted theoretical values from evolutionary models. This discrepancy could be 5--10\% or even more, depending on the model. Up to the beginning of the 21$^{st}$ century, only a few low-mass binary systems with M dwarf companions had measured radii with sufficient accuracy for modelling: 
CM\,Dra \cite{moral2009}, YY\,Gem \cite{torib2002}, CU\,Cnc \cite{ribas2003}, and GU\,Boo \cite{lopez2005}. The situation in 2013 is reviewed in \cite{torres2013}, see citations therein and Fig.\ \ref{fig_book_EBs:mass_rad}). New solutions using more accurate photometric and spectroscopic observations as well as improved models decreased the discrepancy up to 2--3\% for CM\,Dra \cite{feiden2014}. The remaining difference could be caused by uncertain He abundance, for example. The calculation made for CM\,Dra showed that increasing the He abundance by 7\% solves the remaining discrepancy in radii \cite{feiden2014}.   

%{\bf{Milos}}
%We should discuss the impact of stellar activity (spots $+$ flares) on the determination of the
%stellar parameters of each component of the EBs when compared to theoretical predictions.

%
%\subsection%{Activity cycles}
%
%\label{book_EBs:activity_cycles}
%

%{\bf{Ernst}}
%
%Rotation of Praesepe low-mass stars is discussed in \cite{rebull17} and \cite{douglas17a}.
%Pleiades rotations discussed in \cite{rebull16a} and \cite{rebull16b}.
%Hyades rotations discussed in \cite{douglas16a}.

%
%%%%%%%%%%%%%%%%%%%%%%%%%%%%%%%%%%%%%%%%%%%%%%%%
%%%%% Future work on EBs in cluster %%%%%
%%%%%%%%%%%%%%%%%%%%%%%%%%%%%%%%%%%%%%%%%%%%%%%%
%
\section{Future prospects for EBs in clusters}
\label{book_EBs:EBs_future}

The numbers of photometric light curves and transiting systems available in star-forming regions
and open clusters has been overwhelming thanks to the CoRoT mission and the (unexpected) 
advent of the K2 mission focusing on the ecliptic. The future looks very bright too, with the 2-year 
Transiting Exoplanet Survey Satellite (TESS) mission 
\cite{ricker15}\footnote{https://heasarc.gsfc.nasa.gov/docs/tess/} currently in space that will fully 
cover both the southern and northern hemispheres with a cadence of 30\,min. The PLAnetary 
Transits and Oscillations of stars (PLATO) mission \cite{roxburgh07}\footnote{http://sci.esa.int/plato/}
is planned for a launch in 2026 aiming at targeting one million stars with two major objectives:
the discovery of transiting Earth-like planets in the habitable zone of their host star and the study 
of stellar oscillations. However, TESS and PLATO might not contribute too much to the study
of low-mass M dwarf members of star-forming regions and open clusters because they tend
to avoid the ecliptic due to confusion issues and will focus on stars brighter than K2\@.
We encourage the PLATO community to design a specific program focusing on a few clusters
bracketing a large age range for several days to further constrain planetary and stellar evolutionary
models.

The Young Exoplanet Transit Initiative (YETI) is an independent large ground-based involving 
multi-site telescopes and instruments designed to focus on nearby young
regions to look for planetary transits \cite{neuhaueser11}. Only one transiting planetary
candidate (CVSO\,30) has been identified but not yet unambiguously confirmed so far \cite{raetz16}.
As a spin-off result, a few EBs have been discovered in several clusters, including
the NGC\,7243 ($\sim$250 Myr; $\sim$700 pc) and Trumpler\,37 (4 Myr; 840 pc) clusters \cite{errmann13,garai16}. 
The main difficulties of ground-based photometric surveys lies in the length of the nights (8h vs 24h for
space missions) with weather dependent conditions, the limited numbers of dedicated nights (a few
per week/month vs 27/80 days for TESS/K2), and the low precision of single measurements
(at best a few mmag vs less than 1 mmag from space for the same brightness). As a consequence,
transiting exoplanets are tough to identify in active young stars but low-mass EBs should be easier
to spot even for masses below 0.6 M$_{\odot}$.

In Fig.\ \ref{fig_book_EBs:fig_M_R_EBs}, we can clearly see a gap in age between 10 and 120 Myr.
We highlight some dedicated programs to fill up that gap to investigate the evolution of masses and radii 
with age and constrain state-of-the-art isochrones. 
\begin{itemize}
\item First, we should focus on the nearest ($<$200 pc)
open clusters in this age range. The options are limited, resulting only a few regions: IC\,2391 \cite{barrado04b},
IC\,2602 \cite{dobbie10}, IC\,4665 \cite{manzi08}, $\alpha$\,Persei \cite{barrado02a} and NGC\,2451 \cite{huensch04} 
for which $Gaia$ will provide soon revised membership lists with accurate kinematics. 
One of the main issue though is the extension of these clusters in the sky, typically larger than most
optical and infrared detectors. To reach the low-mass M dwarfs in those regions, the infrared camera
VIRCAM on the VISTA telescope \cite{emerson01,dalton06} might be the best option is the 
photometric accuracy can be guaranteed over several nights
or weeks to look for dipping events of a few tens of magnitudes in the lowest mass members.
Another alternative consists in dedicated a month of observing time with the most sensitive cameras
of planet-hunter surveys with a preference to the ones most sensitive to far-red ($\geq$\,750\,nm)
and infrared wavelengths (1$-$2\,$\mu$m) like the Next-Generation Transit Survey (NGTS; \cite{wheatley13}), 
the TRAnsiting Planets and PlanetesImals Small Telescope (TRAPPIST; \cite{gillon11a}),
or the Search for habitable Planets EClipsing ULtra-cOOl Stars (SPECULOOS; \cite{gillon13a})
rather than in the visible such as Trans-atlantic Exoplanet Survey (Tr-ES; \cite{alonso04a}), 
Hungarian Automated Telescope Network (HATNet; \cite{bakos02}), the Wide Angle Search for Planets 
(WASP) North and South \cite{pollacco06}, the XO telescope \cite{mccullough05a},
the Kilodegree Extremely Little Telescope (KELT; \cite{pepper07}).
\item Secondly, members of young nearby moving groups younger than the Pleiades might represent
ideal targets to look for exoplanets and EBs because they share the same age and metallicity\cite{montes01b,zuckerman04,antoja08,gagne14a,pecaut16a}. Several moving groups and associations
have been identified in the Solar neighbourhood (Ursa Major, TW Hya, $\beta$\,Pic, AB Doradus, 
$\eta$\,Chamaeleon, $\epsilon$\,Chamaeleon, Tucana-Horologium) and hundreds of their members 
confirmed spectroscopically \cite{zuckerman04}. The main advantage is the closest distances of these 
members whose kinematics will be refined soon thanks to the releases of the $Gaia$ astrometric datasets.
One of the main drawback, as for all young stars, is the unknown levels of activity that might mimic the
presence of planets. However, the large amplitude and RV modulation of low-mass EBs should be
less affected to the intrinsic stellar activity.
\item A third option is to search for low-mass companions of B- and A-type stars
using X-ray data \cite{Hubrig2007}. Known EBs can be located in a 
colour-magnitude diagram using $Gaia$ data. Systems close to the zero-age-main-sequence
can be easily identified because the contribution of a possible low-mass
companion to the combined colour and absolute magnitude is negligible.
The next steps consists in identifying those systems in the public X-ray 
catalogues such as Chandra \cite{Evans2018}, ROSAT \cite{Boller2016}, 
and XMM-Newton \cite{Traulsen2019}. Normally, more massive stars have only
weak X-ray fluxes compared to their low-mass counterparts \cite{Schroeder2007}.
From their spectrum and the amount of flux in X-rays, a first estimation of the 
astrophysical parameters of the companions can be done \cite{Stelzer2013} to
select systems for further spectroscopic studies.
\end{itemize}
\begin{acknowledgement}
%If you want to include acknowledgments of assistance and the like at the end of an %individual chapter please use the \verb|acknowledgement| environment -- it will %automatically be rendered in line with the preferred layout.
NL supported by the Spanish Ministry of Economy and Competitiveness (MINECO) under the 
grant AYA2015-69350-C3-2-P\@.
%{\bf{EP and MZ ?}}.
%{\bf{acknowledge ERASMUS$+$}}
This research has made use of the Simbad and Vizier databases, operated
at the centre de Donn\'ees Astronomiques de Strasbourg (CDS),
of NASA's Astrophysics Data System Bibliographic Services (ADS), and the WEBDA 
database, operated at the Department of Theoretical Physics and Astrophysics of the 
Masaryk University.
\end{acknowledgement}
%

%
%%%%%%%%%%%%%%%%%%%%%%%%%%%%%%%%%%%%%%%%%%%%%%%%
%%%%% Appendix %%%%%
%%%%%%%%%%%%%%%%%%%%%%%%%%%%%%%%%%%%%%%%%%%%%%%%
%
%\section*{Appendix}
%\addcontentsline{toc}{section}{Appendix}
%
%When placed at the end of a chapter or contribution (as opposed to at the end of the book), the numbering of tables, figures, and %equations in the appendix section continues on from that in the main text. Hence please \textit{do not} use the \verb|appendix| %command when writing an appendix at the end of your chapter or contribution. If there is only one the appendix is designated %``Appendix'', or ``Appendix 1'', or ``Appendix 2'', etc. if there is more than one.

%
%%%%%%%%%%%%%%%%%%%%%%%%%%%%%%%%%%%%%%%%%%%%%%%%
%%%%% Bibliography %%%%%
%%%%%%%%%%%%%%%%%%%%%%%%%%%%%%%%%%%%%%%%%%%%%%%%
%
\bibliographystyle{spphys}
%\bibliographystyle{../../AA/aa}
%\bibliography{ref_chapter_EBs_ERASMUS,mnemonic}
\bibliography{ref_chapter_EBs_ERASMUS_overleaf,mnemonic}

\begin{thebibliography}{100}
\providecommand{\url}[1]{{#1}}
\providecommand{\urlprefix}{URL }
\expandafter\ifx\csname urlstyle\endcsname\relax
  \providecommand{\doi}[1]{DOI \discretionary{}{}{}#1}\else
  \providecommand{\doi}{DOI \discretionary{}{}{}\begingroup
  \urlstyle{rm}\Url}\fi

\bibitem{henry97}
T.~{Henry}, O.~{Franz}, L.~{Wasserman}, G.F. {Benedict}, P.~{Shelus},
  P.~{Ianna}, J.D. {Kirkpatrick}, D.~{McCarthy}, Bull.\@ AAS \textbf{29}, 1278
  (1997)

\bibitem{winters15}
J.G. {Winters}, T.J. {Henry}, J.C. {Lurie}, N.C. {Hambly}, W.C. {Jao}, J.L.
  {Bartlett}, M.R. {Boyd}, S.B. {Dieterich}, C.T. {Finch}, A.D. {Hosey}, P.A.
  {Ianna}, A.R. {Riedel}, K.J. {Slatten}, J.P. {Subasavage}, AJ \textbf{149}, 5
  (2015).
\newblock \doi{10.1088/0004-6256/149/1/5}

\bibitem{kumar63a}
S.S. {Kumar}, E.K.L. {Upton}, AJ \textbf{68}, 76 (1963)

\bibitem{tarter76}
J.C. {Tarter}, Bulletin of the American Astronomical Society \textbf{8}, 517
  (1976)

\bibitem{chabrier00a}
G.~{Chabrier}, I.~{Baraffe}, ARA\&A \textbf{38}, 337 (2000)

\bibitem{southworth05}
J.~{Southworth}, P.F.L. {Maxted}, B.~{Smalley}, A\&A \textbf{429}, 645 (2005).
\newblock \doi{10.1051/0004-6361:20041867}

\bibitem{torres10a}
G.~{Torres}, J.~{Andersen}, A.~{Gim{\'e}nez}, ARA\&A \textbf{18}, 67 (2010).
\newblock \doi{10.1007/s00159-009-0025-1}

\bibitem{popper80a}
D.M. {Popper}, ARA\&A \textbf{18}, 115 (1980).
\newblock \doi{10.1146/annurev.aa.18.090180.000555}

\bibitem{andersen91a}
J.~{Andersen}, ARA\&A \textbf{3}, 91 (1991).
\newblock \doi{10.1007/BF00873538}

\bibitem{southworth14a}
J.~{Southworth}, ArXiv e-prints, arXiv:1411.1219  (2014)

\bibitem{svechnikov04a}
M.A. {Svechnikov}, E.L. {Perevozkina}, VizieR Online Data Catalog
  \textbf{5121}, 5 (2004)

\bibitem{soszynski17a}
I.~{Soszy{\'n}ski}, A.~{Udalski}, M.K. {Szyma{\'n}ski}, {\L}.~{Wyrzykowski},
  K.~{Ulaczyk}, R.~{Poleski}, P.~{Pietrukowicz}, S.~{Koz{\l}owski}, D.M.
  {Skowron}, J.~{Skowron}, P.~{Mr{\'o}z}, M.~{Pawlak}, K.~{Rybicki},
  A.~{Jacyszyn-Dobrzeniecka}, Acta Astronomica \textbf{67}, 297 (2017).
\newblock \doi{10.32023/0001-5237/67.4.1}

\bibitem{helminiak12a}
K.G. {He{\l}miniak}, M.~{Konacki}, M.~{R{\'o}{\.Z}yczka}, J.~{Ka{\l}u{\.Z}ny},
  M.~{Ratajczak}, J.~{Borkowski}, P.~{Sybilski}, M.W. {Muterspaugh}, D.E.
  {Reichart}, K.M. {Ivarsen}, J.B. {Haislip}, J.A. {Crain}, A.C. {Foster}, M.C.
  {Nysewander}, A.P. {LaCluyze}, Monthly Notices of the Royal Astronomical
  Society \textbf{425}, 1245 (2012).
\newblock \doi{10.1111/j.1365-2966.2012.21510.x}

\bibitem{zhang18a}
L.~{Zhang}, H.~{Lu}, X.L. {Han}, L.~{Jiang}, Z.~{Li}, Y.~{Zhang}, Y.~{Hou},
  Y.~{Wang}, Z.~{Cao}, New Astronomy \textbf{61}, 36 (2018).
\newblock \doi{10.1016/j.newast.2017.11.007}

\bibitem{lee17a}
C.H. {Lee}, C.C. {Lin}, Research in Astronomy and Astrophysics \textbf{17}, 15
  (2017).
\newblock \doi{10.1088/1674-4527/17/2/15}

\bibitem{drake09a}
A.J. {Drake}, S.G. {Djorgovski}, A.~{Mahabal}, E.~{Beshore}, S.~{Larson}, M.J.
  {Graham}, R.~{Williams}, E.~{Christensen}, M.~{Catelan}, A.~{Boattini},
  A.~{Gibbs}, R.~{Hill}, R.~{Kowalski}, ApJ \textbf{696}, 870 (2009).
\newblock \doi{10.1088/0004-637X/696/1/870}

\bibitem{mahabal11}
A.A. {Mahabal}, S.G. {Djorgovski}, A.J. {Drake}, C.~{Donalek}, M.J. {Graham},
  R.D. {Williams}, Y.~{Chen}, B.~{Moghaddam}, M.~{Turmon}, E.~{Beshore},
  S.~{Larson}, Bulletin of the Astronomical Society of India \textbf{39}, 387
  (2011)

\bibitem{djorgovski12}
S.G. {Djorgovski}, A.A. {Mahabal}, C.~{Donalek}, M.J. {Graham}, A.J. {Drake},
  B.~{Moghaddam}, M.~{Turmon}, arXiv e-prints  (2012)

\bibitem{tonry18a}
J.L. {Tonry}, L.~{Denneau}, A.N. {Heinze}, B.~{Stalder}, K.W. {Smith}, S.J.
  {Smartt}, C.W. {Stubbs}, H.J. {Weiland}, A.~{Rest}, Publications of the
  Astronomical Society of the Pacific \textbf{130}(6), 064505 (2018).
\newblock \doi{10.1088/1538-3873/aabadf}

\bibitem{borucki10}
W.J. {Borucki}, D.~{Koch}, G.~{Basri}, N.~{Batalha}, T.~{Brown}, D.~{Caldwell},
  J.~{Caldwell}, J.~{Christensen-Dalsgaard}, {53 co-authors}, Science
  \textbf{327}, 977 (2010).
\newblock \doi{10.1126/science.1185402}

\bibitem{howell14}
S.B. {Howell}, C.~{Sobeck}, M.~{Haas}, M.~{Still}, T.~{Barclay}, F.~{Mullally},
  J.~{Troeltzsch}, S.~{Aigrain}, S.T. {Bryson}, D.~{Caldwell}, W.J. {Chaplin},
  W.D. {Cochran}, D.~{Huber}, G.W. {Marcy}, A.~{Miglio}, J.R. {Najita},
  M.~{Smith}, J.D. {Twicken}, J.J. {Fortney}, Publications of the Astronomical
  Society of the Pacific \textbf{126}, 398 (2014).
\newblock \doi{10.1086/676406}

\bibitem{geller09}
A.M. {Geller}, R.D. {Mathieu}, H.C. {Harris}, R.D. {McClure}, AJ \textbf{137},
  3743 (2009).
\newblock \doi{10.1088/0004-6256/137/4/3743}

\bibitem{milliman14}
K.E. {Milliman}, R.D. {Mathieu}, A.M. {Geller}, N.M. {Gosnell}, S.~{Meibom},
  I.~{Platais}, AJ \textbf{148}, 38 (2014).
\newblock \doi{10.1088/0004-6256/148/2/38}

\bibitem{leiner15}
E.M. {Leiner}, R.D. {Mathieu}, N.M. {Gosnell}, A.M. {Geller}, AJ \textbf{150},
  10 (2015).
\newblock \doi{10.1088/0004-6256/150/1/10}

\bibitem{neuhaueser11}
R.~{Neuh{\"a}user}, R.~{Errmann}, A.~{Berndt}, G.~{Maciejewski},
  H.~{Takahashi}, W.P. {Chen}, D.P. {Dimitrov}, T.~{Pribulla}, E.H.
  {Nikogossian}, E.L.N. {Jensen}, L.~{Marschall}, Z.Y. {Wu}, A.~{Kellerer},
  F.M. {Walter}, C.~{Brice{\~n}o}, R.~{Chini}, M.~{Fernandez}, S.~{Raetz},
  G.~{Torres}, D.W. {Latham}, S.N. {Quinn}, A.~{Niedzielski},
  {\L}.~{Bukowiecki}, G.~{Nowak}, T.~{Tomov}, K.~{Tachihara}, S.C.L. {Hu}, L.W.
  {Hung}, D.P. {Kjurkchieva}, V.S. {Radeva}, B.M. {Mihov},
  L.~{Slavcheva-Mihova}, I.N. {Bozhinova}, J.~{Budaj}, M.~{Va{\v n}ko},
  E.~{Kundra}, {\v L}.~{Hamb{\'a}lek}, V.~{Krushevska}, T.~{Movsessian},
  H.~{Harutyunyan}, J.J. {Downes}, J.~{Hernandez}, V.H. {Hoffmeister}, D.H.
  {Cohen}, I.~{Abel}, R.~{Ahmad}, S.~{Chapman}, S.~{Eckert}, J.~{Goodman},
  A.~{Guerard}, H.M. {Kim}, A.~{Koontharana}, J.~{Sokol}, J.~{Trinh},
  Y.~{Wang}, X.~{Zhou}, R.~{Redmer}, U.~{Kramm}, N.~{Nettelmann},
  M.~{Mugrauer}, J.~{Schmidt}, M.~{Moualla}, C.~{Ginski}, C.~{Marka},
  C.~{Adam}, M.~{Seeliger}, S.~{Baar}, T.~{Roell}, T.O.B. {Schmidt},
  L.~{Trepl}, T.~{Eisenbei{\ss}}, S.~{Fiedler}, N.~{Tetzlaff}, E.~{Schmidt},
  M.M. {Hohle}, M.~{Kitze}, N.~{Chakrova}, C.~{Gr{\"a}fe}, K.~{Schreyer}, V.V.
  {Hambaryan}, C.H. {Broeg}, J.~{Koppenhoefer}, A.K. {Pandey}, Astronomische
  Nachrichten \textbf{332}, 547 (2011).
\newblock \doi{10.1002/asna.201111573}

\bibitem{rau09}
A.~{Rau}, S.R. {Kulkarni}, N.M. {Law}, J.S. {Bloom}, D.~{Ciardi}, G.S.
  {Djorgovski}, D.B. {Fox}, A.~{Gal-Yam}, C.C. {Grillmair}, M.M. {Kasliwal},
  P.E. {Nugent}, E.O. {Ofek}, R.M. {Quimby}, W.T. {Reach}, M.~{Shara},
  L.~{Bildsten}, S.B. {Cenko}, A.J. {Drake}, A.V. {Filippenko}, D.J. {Helfand},
  G.~{Helou}, D.A. {Howell}, D.~{Poznanski}, M.~{Sullivan}, Publications of the
  Astronomical Society of the Pacific \textbf{121}, 1334 (2009).
\newblock \doi{10.1086/605911}

\bibitem{law09}
N.M. {Law}, S.R. {Kulkarni}, R.G. {Dekany}, E.O. {Ofek}, R.M. {Quimby}, P.E.
  {Nugent}, J.~{Surace}, C.C. {Grillmair}, J.S. {Bloom}, M.M. {Kasliwal},
  L.~{Bildsten}, T.~{Brown}, S.B. {Cenko}, D.~{Ciardi}, E.~{Croner}, S.G.
  {Djorgovski}, J.~{van Eyken}, A.V. {Filippenko}, D.B. {Fox}, A.~{Gal-Yam},
  D.~{Hale}, N.~{Hamam}, G.~{Helou}, J.~{Henning}, D.A. {Howell},
  J.~{Jacobsen}, R.~{Laher}, S.~{Mattingly}, D.~{McKenna}, A.~{Pickles},
  D.~{Poznanski}, G.~{Rahmer}, A.~{Rau}, W.~{Rosing}, M.~{Shara}, R.~{Smith},
  D.~{Starr}, M.~{Sullivan}, V.~{Velur}, R.~{Walters}, J.~{Zolkower},
  Publications of the Astronomical Society of the Pacific \textbf{121}, 1395
  (2009).
\newblock \doi{10.1086/648598}

\bibitem{RussellMerrill1952}
H.N. {Russell}, J.E. {Merrill}, \emph{{The determination of the elements of
  eclipsing binaries}} (1952)

\bibitem{Zverev1947}
M.S. {Zverev}, B.V. {Kukarkin}, D.Y. {Martynov}, P.P. {Parenago}, N.F.
  {Florya}, V.P. {Tsesevich}, \emph{{Variable stars, Vol. III}} (1947)

\bibitem{Etzel1981}
P.B. {Etzel}, in \emph{Photometric and Spectroscopic Binary Systems} (1981), p.
  111

\bibitem{Budding1973}
E.~{Budding}, Ap\&SS \textbf{22}(1), 87 (1973).
\newblock \doi{10.1007/BF00642825}

\bibitem{BuddingZeilik1987}
E.~{Budding}, M.~{Zeilik}, ApJ \textbf{319}, 827 (1987).
\newblock \doi{10.1086/165500}

\bibitem{wood1971}
D.B. {Wood}, AJ \textbf{76}, 701 (1971).
\newblock \doi{10.1086/111187}

\bibitem{wood1972}
D.B. {Wood}, Technical Report X-110-72-473, Publ. of the Goddard Space Flight
  Center, Greenbelt, MD  (1972)

\bibitem{wood1973}
D.B. {Wood}, PASP \textbf{85}(504), 253 (1973).
\newblock \doi{10.1086/129447}

\bibitem{hill1979}
G.~{Hill}, Publications of the Dominion Astrophysical Observatory Victoria
  \textbf{15}, 298 (1979)

\bibitem{HillRucinski1993}
G.~{Hill}, S.~{Rucinski}, IAU Commission on Close Binary Stars \textbf{21}, 135
  (1993)

\bibitem{rucinski1973}
S.M. {Ruci{\'n}ski}, Acta Astronomica \textbf{23}, 79 (1973)

\bibitem{rucinski1974}
S.M. {Rucinski}, Acta Astronomica \textbf{24}, 119 (1974)

\bibitem{WD1971}
R.E. {Wilson}, E.J. {Devinney}, ApJ \textbf{166}, 605 (1971).
\newblock \doi{10.1086/150986}

\bibitem{WD1972}
R.E. {Wilson}, E.J. {Devinney}, ApJ \textbf{171}, 413 (1972).
\newblock \doi{10.1086/151293}

\bibitem{WD2008}
R.E. {Wilson}, ApJ \textbf{672}(1), 575 (2008).
\newblock \doi{10.1086/523634}

\bibitem{WD2014}
R.E. {Wilson}, W.~{Van Hamme}, ApJ \textbf{780}(2), 151 (2014).
\newblock \doi{10.1088/0004-637X/780/2/151}

\bibitem{PrsaZwitter2005}
A.~{Pr{\v{s}}a}, T.~{Zwitter}, ApJ \textbf{628}(1), 426 (2005).
\newblock \doi{10.1086/430591}

\bibitem{Prsa2016}
A.~{Pr{\v{s}}a}, K.E. {Conroy}, M.~{Horvat}, H.~{Pablo}, A.~{Kochoska},
  S.~{Bloemen}, J.~{Giammarco}, K.M. {Hambleton}, P.~{Degroote}, ApJS
  \textbf{227}(2), 29 (2016).
\newblock \doi{10.3847/1538-4365/227/2/29}

\bibitem{Horvat2018}
M.~{Horvat}, K.E. {Conroy}, H.~{Pablo}, K.M. {Hambleton}, A.~{Kochoska},
  J.~{Giammarco}, A.~{Pr{\v{s}}a}, ApJS \textbf{237}(2), 26 (2018).
\newblock \doi{10.3847/1538-4365/aacd0f}

\bibitem{Prsa2018}
A.~{Pr{\v{s}}a}, \emph{{Modeling and Analysis of Eclipsing Binary Stars; The
  theory and design principles of PHOEBE}} (2018).
\newblock \doi{10.1088/978-0-7503-1287-5}

\bibitem{devor2005}
J.~{Devor}, ApJ \textbf{628}(1), 411 (2005).
\newblock \doi{10.1086/431170}

\bibitem{devor2006}
J.~{Devor}, D.~{Charbonneau}, ApJ \textbf{653}(1), 647 (2006).
\newblock \doi{10.1086/508609}

\bibitem{tamuz2006a}
O.~{Tamuz}, T.~{Mazeh}, P.~{North}, MNRAS \textbf{367}(4), 1521 (2006a).
\newblock \doi{10.1111/j.1365-2966.2006.10049.x}

\bibitem{tamuz2006b}
T.~{Mazeh}, O.~{Tamuz}, P.~{North}, MNRAS \textbf{367}(4), 1531 (2006b).
\newblock \doi{10.1111/j.1365-2966.2006.10050.x}

\bibitem{hadrava2004}
P.~{Hadrava}, Publications of the Astronomical Institute of the Czechoslovak
  Academy of Sciences \textbf{92}, 1 (2004)

\bibitem{pribulla2012}
T.~{Pribulla}, in \emph{From Interacting Binaries to Exoplanets: Essential
  Modeling Tools}, \emph{IAU Symposium}, vol. 282, ed. by M.T. {Richards},
  I.~{Hubeny} (2012), \emph{IAU Symposium}, vol. 282, pp. 279--282.
\newblock \doi{10.1017/S1743921311027566}

\bibitem{Bradstreet2005}
D.H. {Bradstreet}, Society for Astronomical Sciences Annual Symposium
  \textbf{24}, 23 (2005)

\bibitem{terrell2005}
D.~{Terrell}, R.E. {Wilson}, Ap\&SS \textbf{296}(1-4), 221 (2005).
\newblock \doi{10.1007/s10509-005-4449-4}

\bibitem{hambalek2013}
{\v{L}}.~{Hamb{\'a}lek}, T.~{Pribulla}, Contributions of the Astronomical
  Observatory Skalnate Pleso \textbf{43}(1), 27 (2013)

\bibitem{vanhamme1993}
W.~{van Hamme}, AJ \textbf{106}, 2096 (1993).
\newblock \doi{10.1086/116788}

\bibitem{claret2011}
A.~{Claret}, S.~{Bloemen}, A\&A \textbf{529}, A75 (2011).
\newblock \doi{10.1051/0004-6361/201116451}

\bibitem{claret2012}
A.~{Claret}, P.H. {Hauschildt}, S.~{Witte}, A\&A \textbf{546}, A14 (2012).
\newblock \doi{10.1051/0004-6361/201219849}

\bibitem{claret2013}
A.~{Claret}, P.H. {Hauschildt}, S.~{Witte}, A\&A \textbf{552}, A16 (2013).
\newblock \doi{10.1051/0004-6361/201220942}

\bibitem{claret2017}
A.~{Claret}, A\&A \textbf{600}, A30 (2017).
\newblock \doi{10.1051/0004-6361/201629705}

\bibitem{claret2018}
A.~{Claret}, A\&A \textbf{618}, A20 (2018).
\newblock \doi{10.1051/0004-6361/201833060}

\bibitem{neilson2013}
H.R. {Neilson}, J.B. {Lester}, A\&A \textbf{556}, A86 (2013).
\newblock \doi{10.1051/0004-6361/201321888}

\bibitem{lodieu15c}
N.~{Lodieu}, R.~{Alonso}, R.~{Gonz{\'a}lez Hern{\'a}ndez},
  J.~I.~{Sanchis-Ojeda}, N.~{Narita}, Y.~{Kawashima}, K.~{Kawauchi},
  A.~{Su{\'a}rez Mascare{\~n}o}, H.~{Deeg}, {et al.}, A\&A \textbf{584}, A128
  (2015).
\newblock \doi{10.1051/0004-6361/201527464}

\bibitem{stassun06}
K.G. {Stassun}, R.D. {Mathieu}, J.A. {Valenti}, Nature \textbf{440}, 311
  (2006).
\newblock \doi{10.1038/nature04570}

\bibitem{stassun07a}
K.G. {Stassun}, M.~{van den Berg}, E.~{Feigelson}, ApJ \textbf{660}, 704
  (2007).
\newblock \doi{10.1086/513138}

\bibitem{irwin07a}
J.~{Irwin}, S.~{Aigrain}, S.~{Hodgkin}, K.G. {Stassun}, L.~{Hebb}, M.~{Irwin},
  E.~{Moraux}, J.~{Bouvier}, A.~{Alapini}, R.~{Alexander}, D.M. {Bramich},
  J.~{Holtzman}, E.L. {Mart{\'{\i}}n}, M.J. {McCaughrean}, F.~{Pont}, P.E.
  {Verrier}, M.R. {Zapatero Osorio}, Monthly Notices of the Royal Astronomical
  Society \textbf{380}, 541 (2007).
\newblock \doi{10.1111/j.1365-2966.2007.12117.x}

\bibitem{mjm96}
M.J. {McCaughrean}, C.R. {O'Dell}, AJ \textbf{111}, 1977 (1996)

\bibitem{hillenbrand97}
L.A. {Hillenbrand}, AJ \textbf{113}, 1733 (1997)

\bibitem{hillenbrand00}
L.A. {Hillenbrand}, J.M. {Carpenter}, ApJ \textbf{540}, 236 (2000)

\bibitem{feigelson03}
E.D. {Feigelson}, J.A. {Gaffney}, G.~{Garmire}, L.A. {Hillenbrand},
  L.~{Townsley}, ApJ \textbf{584}, 911 (2003)

\bibitem{hillenbrand13}
L.A. {Hillenbrand}, A.S. {Hoffer}, G.J. {Herczeg}, AJ \textbf{146}, 85 (2013).
\newblock \doi{10.1088/0004-6256/146/4/85}

\bibitem{ingraham14}
P.~{Ingraham}, L.~{Albert}, R.~{Doyon}, E.~{Artigau}, ApJ \textbf{782}, 8
  (2014).
\newblock \doi{10.1088/0004-637X/782/1/8}

\bibitem{stassun08a}
K.G. {Stassun}, R.D. {Mathieu}, P.A. {Cargile}, A.N. {Aarnio}, E.~{Stempels},
  A.~{Geller}, Nature \textbf{453}, 1079 (2008).
\newblock \doi{10.1038/nature07069}

\bibitem{cargile08}
P.A. {Cargile}, K.G. {Stassun}, R.D. {Mathieu}, ApJ \textbf{674}, 329 (2008).
\newblock \doi{10.1086/524346}

\bibitem{gomez_maqueo12}
Y.~{G{\'o}mez Maqueo Chew}, K.G. {Stassun}, A.~{Pr{\v s}a}, E.~{Stempels},
  L.~{Hebb}, R.~{Barnes}, R.~{Heller}, R.D. {Mathieu}, ApJ \textbf{745}, 58
  (2012).
\newblock \doi{10.1088/0004-637X/745/1/58}

\bibitem{morales_calderon12}
M.~{Morales-Calder{\'o}n}, J.R. {Stauffer}, K.G. {Stassun}, F.J. {Vrba},
  L.~{Prato}, L.A. {Hillenbrand}, S.~{Terebey}, K.R. {Covey}, L.M. {Rebull},
  D.M. {Terndrup}, R.~{Gutermuth}, I.~{Song}, P.~{Plavchan}, J.M. {Carpenter},
  F.~{Marchis}, E.V. {Garc{\'{\i}}a}, S.~{Margheim}, K.L. {Luhman},
  J.~{Angione}, J.M. {Irwin}, ApJ \textbf{753}, 149 (2012).
\newblock \doi{10.1088/0004-637X/753/2/149}

\bibitem{preibisch99}
T.~{Preibisch}, H.~{Zinnecker}, AJ \textbf{117}, 2381 (1999).
\newblock \doi{10.1086/300842}

\bibitem{preibisch01}
T.~{Preibisch}, E.~{Guenther}, H.~{Zinnecker}, AJ \textbf{121}, 1040 (2001).
\newblock \doi{10.1086/318774}

\bibitem{slesnick08}
C.L. {Slesnick}, L.A. {Hillenbrand}, J.M. {Carpenter}, ApJ \textbf{688}, 377
  (2008).
\newblock \doi{10.1086/592265}

\bibitem{pecaut12}
M.J. {Pecaut}, E.E. {Mamajek}, E.J. {Bubar}, ApJ \textbf{746}, 154 (2012).
\newblock \doi{10.1088/0004-637X/746/2/154}

\bibitem{song12}
I.~{Song}, B.~{Zuckerman}, M.S. {Bessell}, AJ \textbf{144}, 8 (2012).
\newblock \doi{10.1088/0004-6256/144/1/8}

\bibitem{pecaut16a}
M.J. {Pecaut}, in \emph{IAU Symposium}, \emph{IAU Symposium}, vol. 314, ed. by
  J.H. {Kastner}, B.~{Stelzer}, S.A. {Metchev} (2016), \emph{IAU Symposium},
  vol. 314, pp. 85--90.
\newblock \doi{10.1017/S1743921315006079}

\bibitem{rizzuto16}
A.C. {Rizzuto}, M.J. {Ireland}, T.J. {Dupuy}, A.L. {Kraus}, ApJ \textbf{817},
  164 (2016).
\newblock \doi{10.3847/0004-637X/817/2/164}

\bibitem{david19a}
T.J. {David}, L.A. {Hillenbrand}, E.~{Gillen}, A.M. {Cody}, S.B. {Howell}, H.T.
  {Isaacson}, J.H. {Livingston}, ApJ \textbf{872}, 161 (2019).
\newblock \doi{10.3847/1538-4357/aafe09}

\bibitem{slesnick06}
C.L. {Slesnick}, J.M. {Carpenter}, L.A. {Hillenbrand}, AJ \textbf{131}, 3016
  (2006).
\newblock \doi{10.1086/503560}

\bibitem{lodieu13c}
N.~{Lodieu}, Monthly Notices of the Royal Astronomical Society \textbf{431},
  3222 (2013).
\newblock \doi{10.1093/{Monthly Notices of the Royal Astronomical
  Society}/stt402}

\bibitem{dawson13}
P.~{Dawson}, A.~{Scholz}, T.P. {Ray}, K.A. {Marsh}, K.~{Wood}, A.~{Natta},
  D.~{Padgett}, M.E. {Ressler}, Monthly Notices of the Royal Astronomical
  Society \textbf{429}, 903 (2013).
\newblock \doi{10.1093/{Monthly Notices of the Royal Astronomical
  Society}/sts386}

\bibitem{ardila00}
D.~{Ardila}, E.~{Mart{\'{\i}}n}, G.~{Basri}, AJ \textbf{120}, 479 (2000)

\bibitem{reiners05a}
A.~{Reiners}, G.~{Basri}, S.~{Mohanty}, ApJ \textbf{634}, 1346 (2005).
\newblock \doi{10.1086/432878}

\bibitem{kraus15a}
A.L. {Kraus}, A.M. {Cody}, K.R. {Covey}, A.C. {Rizzuto}, A.W. {Mann}, M.J.
  {Ireland}, ApJ \textbf{807}, 3 (2015).
\newblock \doi{10.1088/0004-637X/807/1/3}

\bibitem{david16a}
T.J. {David}, L.A. {Hillenbrand}, A.M. {Cody}, J.M. {Carpenter}, A.W. {Howard},
  ApJ \textbf{816}, 21 (2016).
\newblock \doi{10.3847/0004-637X/816/1/21}

\bibitem{alonso15a}
R.~{Alonso}, H.J. {Deeg}, S.~{Hoyer}, N.~{Lodieu}, E.~{Palle},
  R.~{Sanchis-Ojeda}, A\&A \textbf{584}, L8 (2015).
\newblock \doi{10.1051/0004-6361/201527109}

\bibitem{stauffer98}
J.R. {Stauffer}, G.~{Schultz}, J.D. {Kirkpatrick}, ApJL \textbf{499}, 199
  (1998)

\bibitem{barrado04b}
D.~{Barrado y Navascu{\' e}s}, J.R. {Stauffer}, R.~{Jayawardhana}, ApJ
  \textbf{614}, 386 (2004)

\bibitem{mazzei89}
P.~{Mazzei}, L.~{Pigatto}, A\&A \textbf{213}, L1 (1989)

\bibitem{dahm15}
S.E. {Dahm}, ApJ \textbf{813}, 108 (2015).
\newblock \doi{10.1088/0004-637X/813/2/108}

\bibitem{gossage18}
S.~{Gossage}, C.~{Conroy}, A.~{Dotter}, J.~{Choi}, P.~{Rosenfield},
  P.~{Cargile}, A.~{Dolphin}, ApJ \textbf{863}, 67 (2018).
\newblock \doi{10.3847/1538-4357/aad0a0}

\bibitem{lodieu19b}
N.~{Lodieu}, A.~{Perez-Garrido}, R.L. {Smart}, R.~{Silvotti}, A\&A  (2019)

\bibitem{maeder81}
A.~{Maeder}, J.C. {Mermilliod}, A\&A \textbf{93}, 136 (1981)

\bibitem{mermilliod81}
J.C. {Mermilliod}, A\&A \textbf{97}, 235 (1981)

\bibitem{mazzei88}
P.~{Mazzei}, L.~{Pigatto}, A\&A \textbf{193}, 148 (1988)

\bibitem{lebreton01}
Y.~{Lebreton}, J.~{Fernandes}, T.~{Lejeune}, A\&A \textbf{374}, 540 (2001).
\newblock \doi{10.1051/0004-6361:20010757}

\bibitem{deGennaro09}
S.~{De Gennaro}, T.~{von Hippel}, W.H. {Jefferys}, N.~{Stein}, D.~{van Dyk},
  E.~{Jeffery}, ApJ \textbf{696}, 12 (2009).
\newblock \doi{10.1088/0004-637X/696/1/12}

\bibitem{brandt15a}
T.D. {Brandt}, C.X. {Huang}, ApJ \textbf{807}, 58 (2015).
\newblock \doi{10.1088/0004-637X/807/1/58}

\bibitem{martin18a}
E.L. {Mart{\'{\i}}n}, N.~{Lodieu}, Y.~{Pavlenko}, V.J.S. {B{\'e}jar}, ApJ
  \textbf{856}, 40 (2018).
\newblock \doi{10.3847/1538-4357/aaaeb8}

\bibitem{lodieu18b}
N.~{Lodieu}, R.~{Rebolo}, A.~{P{\'e}rez-Garrido}, A\&A \textbf{615}, L12
  (2018).
\newblock \doi{10.1051/0004-6361/201832748}

\bibitem{vandenberg84}
D.A. {Vandenberg}, T.J. {Bridges}, ApJ \textbf{278}, 679 (1984).
\newblock \doi{10.1086/161836}

\bibitem{delorme11}
P.~{Delorme}, A.~{Collier Cameron}, L.~{Hebb}, J.~{Rostron}, T.A. {Lister},
  A.J. {Norton}, D.~{Pollacco}, R.G. {West}, Monthly Notices of the Royal
  Astronomical Society \textbf{413}, 2218 (2011).
\newblock \doi{10.1111/j.1365-2966.2011.18299.x}

\bibitem{salaris04}
M.~{Salaris}, A.~{Weiss}, S.M. {Percival}, A\&A \textbf{414}, 163 (2004).
\newblock \doi{10.1051/0004-6361:20031578}

\bibitem{bonatto04a}
C.~{Bonatto}, E.~{Bica}, L.~{Girardi}, A\&A \textbf{415}, 571 (2004).
\newblock \doi{10.1051/0004-6361:20034638}

\bibitem{babusiaux18}
{Gaia Collaboration}, C.~{Babusiaux}, F.~{van Leeuwen}, M.A. {Barstow},
  C.~{Jordi}, A.~{Vallenari}, D.~{Bossini}, A.~{Bressan}, T.~{Cantat-Gaudin},
  M.~{van Leeuwen}, et~al., A\&A \textbf{616}, A10 (2018).
\newblock \doi{10.1051/0004-6361/201832843}

\bibitem{boesgaard90}
A.M. {Boesgaard}, E.D. {Friel}, ApJ \textbf{351}, 467 (1990)

\bibitem{cayreldestrobel97}
G.~{Cayrel de Strobel}, F.~{Crifo}, Y.~{Lebreton}, in \emph{Hipparcos - Venice
  '97}, \emph{ESA Special Publication}, vol. 402, ed. by R.M. {Bonnet},
  E.~{H{\o}g}, P.L. {Bernacca}, L.~{Emiliani}, A.~{Blaauw}, C.~{Turon},
  J.~{Kovalevsky}, L.~{Lindegren}, H.~{Hassan}, M.~{Bouffard}, B.~{Strim},
  D.~{Heger}, M.A.C. {Perryman}, L.~{Woltjer} (1997), \emph{ESA Special
  Publication}, vol. 402, pp. 687--688

\bibitem{grenon00}
M.~{Grenon}, in \emph{IAU Joint Discussion}, \emph{IAU Joint Discussion},
  vol.~13 (2000), \emph{IAU Joint Discussion}, vol.~13

\bibitem{david15b}
T.J. {David}, J.~{Stauffer}, L.A. {Hillenbrand}, A.M. {Cody}, K.~{Conroy}, K.G.
  {Stassun}, B.~{Pope}, S.~{Aigrain}, E.~{Gillen}, A.~{Collier Cameron},
  D.~{Barrado}, L.M. {Rebull}, H.~{Isaacson}, G.W. {Marcy}, C.~{Zhang}, R.L.
  {Riddle}, C.~{Ziegler}, N.M. {Law}, C.~{Baranec}, ApJ \textbf{814}, 62
  (2015).
\newblock \doi{10.1088/0004-637X/814/1/62}

\bibitem{david16c}
T.J. {David}, K.E. {Conroy}, L.A. {Hillenbrand}, K.G. {Stassun}, J.~{Stauffer},
  L.M. {Rebull}, A.M. {Cody}, H.~{Isaacson}, A.W. {Howard}, S.~{Aigrain}, AJ
  \textbf{151}, 112 (2016).
\newblock \doi{10.3847/0004-6256/151/5/112}

\bibitem{hebb06}
L.~{Hebb}, R.F.G. {Wyse}, G.~{Gilmore}, J.~{Holtzman}, AJ \textbf{131}, 555
  (2006).
\newblock \doi{10.1086/497971}

\bibitem{dias02a}
W.S. {Dias}, B.S. {Alessi}, A.~{Moitinho}, J.R.D. {L{\'e}pine}, A\&A
  \textbf{389}, 871 (2002).
\newblock \doi{10.1051/0004-6361:20020668}

\bibitem{turner92a}
D.G. {Turner}, AJ \textbf{104}, 1865 (1992).
\newblock \doi{10.1086/116363}

\bibitem{kraus17a}
A.L. {Kraus}, G.J. {Herczeg}, A.C. {Rizzuto}, A.W. {Mann}, C.L. {Slesnick},
  J.M. {Carpenter}, L.A. {Hillenbrand}, E.E. {Mamajek}, ApJ \textbf{838}, 150
  (2017).
\newblock \doi{10.3847/1538-4357/aa62a0}

\bibitem{gillen17a}
E.~{Gillen}, L.A. {Hillenbrand}, T.J. {David}, S.~{Aigrain}, L.~{Rebull},
  J.~{Stauffer}, A.M. {Cody}, D.~{Queloz}, ApJ \textbf{849}, 11 (2017).
\newblock \doi{10.3847/1538-4357/aa84b3}

\bibitem{adams02a}
J.D. {Adams}, J.R. {Stauffer}, M.F. {Skrutskie}, D.G. {Monet}, S.F. {Portegies
  Zwart}, K.A. {Janes}, C.A. {Beichman}, AJ \textbf{124}, 1570 (2002).
\newblock \doi{10.1086/342016}

\bibitem{kraus07d}
A.L. {Kraus}, L.A. {Hillenbrand}, AJ \textbf{134}, 2340 (2007).
\newblock \doi{10.1086/522831}

\bibitem{baker10}
D.E.A. {Baker}, R.F. {Jameson}, S.L. {Casewell}, N.~{Deacon}, N.~{Lodieu},
  N.~{Hambly}, Monthly Notices of the Royal Astronomical Society \textbf{408},
  2457 (2010).
\newblock \doi{10.1111/j.1365-2966.2010.17302.x}

\bibitem{boudreault12}
S.~{Boudreault}, N.~{Lodieu}, N.R. {Deacon}, N.C. {Hambly}, Monthly Notices of
  the Royal Astronomical Society \textbf{426}, 3419 (2012).
\newblock \doi{10.1111/j.1365-2966.2012.21854.x}

\bibitem{khalaj13}
P.~{Khalaj}, H.~{Baumgardt}, Monthly Notices of the Royal Astronomical Society
  \textbf{434}, 3236 (2013).
\newblock \doi{10.1093/{Monthly Notices of the Royal Astronomical
  Society}/stt1239}

\bibitem{wang14a}
P.F. {Wang}, W.P. {Chen}, C.C. {Lin}, A.K. {Pandey}, C.K. {Huang}, N.~{Panwar},
  C.H. {Lee}, M.F. {Tsai}, C.H. {Tang}, B.~{Goldman}, W.S. {Burgett}, K.C.
  {Chambers}, P.W. {Draper}, H.~{Flewelling}, T.~{Grav}, J.N. {Heasley}, K.W.
  {Hodapp}, M.E. {Huber}, R.~{Jedicke}, N.~{Kaiser}, R.P. {Kudritzki}, G.A.
  {Luppino}, R.H. {Lupton}, E.A. {Magnier}, N.~{Metcalfe}, D.G. {Monet}, J.S.
  {Morgan}, P.M. {Onaka}, P.A. {Price}, C.W. {Stubbs}, W.~{Sweeney}, J.L.
  {Tonry}, R.J. {Wainscoat}, C.~{Waters}, ApJ \textbf{784}, 57 (2014).
\newblock \doi{10.1088/0004-637X/784/1/57}

\bibitem{mann17a}
A.W. {Mann}, E.~{Gaidos}, A.~{Vanderburg}, A.C. {Rizzuto}, M.~{Ansdell}, J.V.
  {Medina}, G.N. {Mace}, A.L. {Kraus}, K.R. {Sokal}, AJ \textbf{153}, 64
  (2017).
\newblock \doi{10.1088/1361-6528/aa5276}

\bibitem{vanAltena66a}
W.F. {van Altena}, AJ \textbf{71}, 482 (1966).
\newblock \doi{10.1086/109952}

\bibitem{hanson75}
R.B. {Hanson}, AJ \textbf{80}, 379 (1975).
\newblock \doi{10.1086/111753}

\bibitem{mann16a}
A.W. {Mann}, E.~{Gaidos}, G.N. {Mace}, M.C. {Johnson}, B.P. {Bowler},
  D.~{LaCourse}, T.L. {Jacobs}, A.~{Vanderburg}, A.L. {Kraus}, K.F. {Kaplan},
  D.T. {Jaffe}, ApJ \textbf{818}, 46 (2016).
\newblock \doi{10.3847/0004-637X/818/1/46}

\bibitem{lodieu19a}
N.~{Lodieu}, R.L. {Smart}, A.~{Perez-Garrido}, R.~{Silvotti}, A\&A  (2019)

\bibitem{Baumgardt2018}
H.~{Baumgardt}, M.~{Hilker}, Monthly Notices of the Royal Astronomical Society
  \textbf{478}, 1520 (2018).
\newblock \doi{10.1093/mnras/sty1057}

\bibitem{Rossi2015}
L.J. {Rossi}, S.~{Ortolani}, B.~{Barbuy}, E.~{Bica}, A.~{Bonfanti}, Monthly
  Notices of the Royal Astronomical Society \textbf{450}, 3270 (2015).
\newblock \doi{10.1093/mnras/stv748}

\bibitem{Bica2006}
E.~{Bica}, C.~{Bonatto}, B.~{Barbuy}, S.~{Ortolani}, A\&A \textbf{450}, 105
  (2006).
\newblock \doi{10.1051/0004-6361:20054351}

\bibitem{Piotto2007}
G.~{Piotto}, L.R. {Bedin}, J.~{Anderson}, I.R. {King}, S.~{Cassisi}, A.P.
  {Milone}, S.~{Villanova}, A.~{Pietrinferni}, A.~{Renzini}, ApJL \textbf{661},
  L53 (2007).
\newblock \doi{10.1086/518503}

\bibitem{Cassisi2017}
S.~{Cassisi}, M.~{Salaris}, A.~{Pietrinferni}, D.~{Hyder}, Monthly Notices of
  the Royal Astronomical Society \textbf{464}, 2341 (2017).
\newblock \doi{10.1093/mnras/stw2579}

\bibitem{Kaluzny2007}
J.~{Kaluzny}, S.M. {Rucinski}, I.B. {Thompson}, W.~{Pych}, W.~{Krzeminski}, AJ
  \textbf{133}, 2457 (2007).
\newblock \doi{10.1086/516637}

\bibitem{Kaluzny2015}
J.~{Kaluzny}, I.B. {Thompson}, A.~{Dotter}, M.~{Rozyczka},
  A.~{Schwarzenberg-Czerny}, G.S. {Burley}, B.~{Mazur}, S.M. {Rucinski}, AJ
  \textbf{150}, 155 (2015).
\newblock \doi{10.1088/0004-6256/150/5/155}

\bibitem{Sandage1953}
A.R. {Sandage}, AJ \textbf{58}, 61 (1953).
\newblock \doi{10.1086/106822}

\bibitem{Sollima2007}
A.~{Sollima}, G.~{Beccari}, F.R. {Ferraro}, F.~{Fusi Pecci}, A.~{Sarajedini},
  Monthly Notices of the Royal Astronomical Society \textbf{380}, 781 (2007).
\newblock \doi{10.1111/j.1365-2966.2007.12116.x}

\bibitem{Ji2013}
J.~{Ji}, J.N. {Bregman}, ApJ \textbf{768}, 158 (2013).
\newblock \doi{10.1088/0004-637X/768/2/158}

\bibitem{Lucatello2015}
S.~{Lucatello}, A.~{Sollima}, R.~{Gratton}, E.~{Vesperini}, V.~{D'Orazi},
  E.~{Carretta}, A.~{Bragaglia}, A\&A \textbf{584}, A52 (2015).
\newblock \doi{10.1051/0004-6361/201526957}

\bibitem{Milone2016}
A.P. {Milone}, A.F. {Marino}, L.R. {Bedin}, A.~{Dotter}, H.~{Jerjen}, D.~{Kim},
  D.~{Nardiello}, G.~{Piotto}, J.~{Cong}, Monthly Notices of the Royal
  Astronomical Society \textbf{455}, 3009 (2016).
\newblock \doi{10.1093/mnras/stv2415}

\bibitem{Clement2001}
C.M. {Clement}, A.~{Muzzin}, Q.~{Dufton}, T.~{Ponnampalam}, J.~{Wang},
  J.~{Burford}, A.~{Richardson}, T.~{Rosebery}, J.~{Rowe}, H.S. {Hogg}, AJ
  \textbf{122}, 2587 (2001).
\newblock \doi{10.1086/323719}

\bibitem{Percy2011}
J.R. {Percy}, \emph{{Understanding Variable Stars}} (2011)

\bibitem{Deras2019}
D.~{Deras}, A.~{Arellano Ferro}, C.~{L{\'a}zaro}, I.H. {Bustos Fierro}, J.H.
  {Calder{\'o}n}, S.~{Muneer}, S.~{Giridhar}, Monthly Notices of the Royal
  Astronomical Society \textbf{486}, 2791 (2019).
\newblock \doi{10.1093/mnras/stz642}

\bibitem{Yepez2018}
M.A. {Yepez}, A.~{Arellano Ferro}, S.~{Muneer}, S.~{Giridhar}, RMxAA
  \textbf{54}, 15 (2018)

\bibitem{Rozyczka2017}
M.~{Rozyczka}, I.B. {Thompson}, W.~{Pych}, W.~{Narloch}, R.~{Poleski},
  A.~{Schwarzenberg-Czerny}, Acta Astronomica \textbf{67}, 203 (2017).
\newblock \doi{10.32023/0001-5237/67.3.1}

\bibitem{Tsapras2017}
Y.~{Tsapras}, A.~{Arellano Ferro}, D.M. {Bramich}, R.F. {Jaimes}, N.~{Kains},
  R.~{Street}, M.~{Hundertmark}, K.~{Horne}, M.~{Dominik}, C.~{Snodgrass},
  Monthly Notices of the Royal Astronomical Society \textbf{465}, 2489 (2017).
\newblock \doi{10.1093/mnras/stw2773}

\bibitem{Lee2016}
D.J. {Lee}, J.R. {Koo}, K.~{Hong}, S.L. {Kim}, J.W. {Lee}, C.U. {Lee}, Y.B.
  {Jeon}, Y.H. {Kim}, B.~{Lim}, Y.H. {Ryu}, S.M. {Cha}, Y.~{Lee}, D.J. {Kim},
  B.G. {Park}, C.H. {Kim}, Journal of Korean Astronomical Society \textbf{49},
  295 (2016).
\newblock \doi{10.5303/JKAS.2016.49.6.295}

\bibitem{fischer92}
D.A. {Fischer}, G.W. {Marcy}, ApJ \textbf{396}, 178 (1992)

\bibitem{baroch18}
D.~{Baroch}, J.C. {Morales}, I.~{Ribas}, L.~{Tal-Or}, M.~{Zechmeister},
  A.~{Reiners}, J.A. {Caballero}, A.~{Quirrenbach}, P.J. {Amado},
  S.~{Dreizler}, S.~{Lalitha}, S.V. {Jeffers}, M.~{Lafarga}, V.J.S.
  {B{\'e}jar}, J.~{Colom{\'e}}, M.~{Cort{\'e}s-Contreras},
  E.~{D{\'{\i}}ez-Alonso}, D.~{Galad{\'{\i}}-Enr{\'{\i}}quez}, E.W. {Guenther},
  H.J. {Hagen}, T.~{Henning}, E.~{Herrero}, M.~{K{\"u}rster}, D.~{Montes},
  E.~{Nagel}, V.M. {Passegger}, M.~{Perger}, A.~{Rosich}, A.~{Schweitzer},
  W.~{Seifert}, A\&A \textbf{619}, A32 (2018).
\newblock \doi{10.1051/0004-6361/201833440}

\bibitem{clark12a}
B.M. {Clark}, C.H. {Blake}, G.R. {Knapp}, ApJ \textbf{744}, 119 (2012).
\newblock \doi{10.1088/0004-637X/744/2/119}

\bibitem{basri06}
G.~{Basri}, A.~{Reiners}, AJ \textbf{132}, 663 (2006).
\newblock \doi{10.1086/505198}

\bibitem{blake10a}
C.H. {Blake}, D.~{Charbonneau}, R.J. {White}, ApJ \textbf{723}(1), 684 (2010).
\newblock \doi{10.1088/0004-637X/723/1/684}

\bibitem{shan15a}
Y.~{Shan}, J.A. {Johnson}, T.D. {Morton}, ApJ \textbf{813}, 75 (2015).
\newblock \doi{10.1088/0004-637X/813/1/75}

\bibitem{rebull16a}
L.M. {Rebull}, J.R. {Stauffer}, J.~{Bouvier}, A.M. {Cody}, L.A. {Hillenbrand},
  D.R. {Soderblom}, J.~{Valenti}, D.~{Barrado}, H.~{Bouy}, D.~{Ciardi},
  M.~{Pinsonneault}, K.~{Stassun}, G.~{Micela}, S.~{Aigrain}, F.~{Vrba},
  G.~{Somers}, E.~{Gillen}, A.~{Collier Cameron}, AJ \textbf{152}, 114 (2016).
\newblock \doi{10.3847/0004-6256/152/5/114}

\bibitem{stauffer16a}
J.~{Stauffer}, L.~{Rebull}, J.~{Bouvier}, L.A. {Hillenbrand},
  A.~{Collier-Cameron}, M.~{Pinsonneault}, S.~{Aigrain}, D.~{Barrado},
  H.~{Bouy}, D.~{Ciardi}, A.M. {Cody}, T.~{David}, G.~{Micela}, D.~{Soderblom},
  G.~{Somers}, K.G. {Stassun}, J.~{Valenti}, F.J. {Vrba}, AJ \textbf{152}, 115
  (2016).
\newblock \doi{10.3847/0004-6256/152/5/115}

\bibitem{rebull17}
L.M. {Rebull}, J.R. {Stauffer}, L.A. {Hillenbrand}, A.M. {Cody}, J.~{Bouvier},
  D.R. {Soderblom}, M.~{Pinsonneault}, L.~{Hebb}, ApJ  (2017)

\bibitem{rebull18}
L.M. {Rebull}, J.R. {Stauffer}, A.M. {Cody}, L.A. {Hillenbrand}, T.J. {David},
  M.~{Pinsonneault}, AJ \textbf{155}, 196 (2018).
\newblock \doi{10.3847/1538-3881/aab605}

\bibitem{stassun07b}
K.G. {Stassun}, R.D. {Mathieu}, J.A. {Valenti}, ApJ \textbf{664}, 1154 (2007).
\newblock \doi{10.1086/519231}

\bibitem{burrows93}
A.~{Burrows}, J.~{Liebert}, Reviews of Modern Physics \textbf{65}, 301 (1993)

\bibitem{baraffe98}
I.~{Baraffe}, G.~{Chabrier}, F.~{Allard}, P.H. {Hauschildt}, A\&A \textbf{337},
  403 (1998)

\bibitem{baraffe15}
I.~{Baraffe}, D.~{Homeier}, F.~{Allard}, G.~{Chabrier}, A\&A \textbf{577}, A42
  (2015).
\newblock \doi{10.1051/0004-6361/201425481}

\bibitem{gillen14}
E.~{Gillen}, S.~{Aigrain}, A.~{McQuillan}, J.~{Bouvier}, S.~{Hodgkin}, S.H.P.
  {Alencar}, C.~{Terquem}, J.~{Southworth}, N.P. {Gibson}, A.~{Cody},
  M.~{Lendl}, M.~{Morales-Calder{\'o}n}, F.~{Favata}, J.~{Stauffer},
  G.~{Micela}, A\&A \textbf{562}, A50 (2014).
\newblock \doi{10.1051/0004-6361/201322493}

\bibitem{anderson11a}
D.R. {Anderson}, A.~{Collier Cameron}, C.~{Hellier}, M.~{Lendl}, P.F.L.
  {Maxted}, D.~{Pollacco}, D.~{Queloz}, B.~{Smalley}, A.M.S. {Smith},
  I.~{Todd}, A.H.M.J. {Triaud}, R.G. {West}, S.C.C. {Barros}, B.~{Enoch},
  M.~{Gillon}, T.A. {Lister}, F.~{Pepe}, D.~{S{\'e}gransan}, R.A. {Street},
  S.~{Udry}, ApJL \textbf{726}, L19 (2011).
\newblock \doi{10.1088/2041-8205/726/2/L19}

\bibitem{diaz14a}
R.F. {D{\'{\i}}az}, G.~{Montagnier}, J.~{Leconte}, A.S. {Bonomo}, M.~{Deleuil},
  J.M. {Almenara}, S.C.C. {Barros}, F.~{Bouchy}, G.~{Bruno}, C.~{Damiani},
  G.~{H{\'e}brard}, C.~{Moutou}, A.~{Santerne}, A\&A \textbf{572}, A109 (2014).
\newblock \doi{10.1051/0004-6361/201424406}

\bibitem{bonomo15}
A.S. {Bonomo}, A.~{Sozzetti}, A.~{Santerne}, M.~{Deleuil}, J.M. {Almenara},
  G.~{Bruno}, R.F. {D{\'{\i}}az}, G.~{H{\'e}brard}, C.~{Moutou}, A\&A
  \textbf{575}, A85 (2015).
\newblock \doi{10.1051/0004-6361/201323042}

\bibitem{bouchy11a}
F.~{Bouchy}, M.~{Deleuil}, T.~{Guillot}, S.~{Aigrain}, L.~{Carone}, W.D.
  {Cochran}, J.M. {Almenara}, R.~{Alonso}, M.~{Auvergne}, A.~{Baglin},
  P.~{Barge}, A.S. {Bonomo}, P.~{Bord{\'e}}, S.~{Csizmadia}, K.~{de Bondt},
  H.J. {Deeg}, R.F. {D{\'{\i}}az}, R.~{Dvorak}, M.~{Endl}, A.~{Erikson},
  S.~{Ferraz-Mello}, M.~{Fridlund}, D.~{Gandolfi}, J.C. {Gazzano}, N.~{Gibson},
  M.~{Gillon}, E.~{Guenther}, A.~{Hatzes}, M.~{Havel}, G.~{H{\'e}brard},
  L.~{Jorda}, A.~{L{\'e}ger}, C.~{Lovis}, A.~{Llebaria}, H.~{Lammer}, P.J.
  {MacQueen}, T.~{Mazeh}, C.~{Moutou}, A.~{Ofir}, M.~{Ollivier},
  H.~{Parviainen}, M.~{P{\"a}tzold}, D.~{Queloz}, H.~{Rauer}, D.~{Rouan},
  A.~{Santerne}, J.~{Schneider}, B.~{Tingley}, G.~{Wuchterl}, A\&A
  \textbf{525}, A68 (2011).
\newblock \doi{10.1051/0004-6361/201015276}

\bibitem{csizmadia15}
S.~{Csizmadia}, A.~{Hatzes}, D.~{Gandolfi}, M.~{Deleuil}, F.~{Bouchy},
  M.~{Fridlund}, L.~{Szabados}, H.~{Parviainen}, J.~{Cabrera}, S.~{Aigrain},
  R.~{Alonso}, J.M. {Almenara}, A.~{Baglin}, P.~{Bord{\'e}}, A.S. {Bonomo},
  H.J. {Deeg}, R.F. {D{\'{\i}}az}, A.~{Erikson}, S.~{Ferraz-Mello}, M.~{Tadeu
  dos Santos}, E.W. {Guenther}, T.~{Guillot}, S.~{Grziwa}, G.~{H{\'e}brard},
  P.~{Klagyivik}, M.~{Ollivier}, M.~{P{\"a}tzold}, H.~{Rauer}, D.~{Rouan},
  A.~{Santerne}, J.~{Schneider}, T.~{Mazeh}, G.~{Wuchterl}, S.~{Carpano},
  A.~{Ofir}, A\&A \textbf{584}, A13 (2015).
\newblock \doi{10.1051/0004-6361/201526763}

\bibitem{siverd12}
R.J. {Siverd}, T.G. {Beatty}, J.~{Pepper}, J.D. {Eastman}, K.~{Collins},
  A.~{Bieryla}, D.W. {Latham}, L.A. {Buchhave}, E.L.N. {Jensen}, J.R. {Crepp},
  R.~{Street}, K.G. {Stassun}, B.S. {Gaudi}, P.~{Berlind}, M.L. {Calkins}, D.L.
  {DePoy}, G.A. {Esquerdo}, B.J. {Fulton}, G.~{F{\H u}r{\'e}sz}, J.C. {Geary},
  A.~{Gould}, L.~{Hebb}, J.F. {Kielkopf}, J.L. {Marshall}, R.~{Pogge}, K.Z.
  {Stanek}, R.P. {Stefanik}, A.H. {Szentgyorgyi}, M.~{Trueblood},
  P.~{Trueblood}, A.M. {Stutz}, J.L. {van Saders}, ApJ \textbf{761}, 123
  (2012).
\newblock \doi{10.1088/0004-637X/761/2/123}

\bibitem{mamajek08}
E.E. {Mamajek}, L.A. {Hillenbrand}, ApJ \textbf{687}, 1264 (2008).
\newblock \doi{10.1086/591785}

\bibitem{irwin10}
J.~{Irwin}, L.~{Buchhave}, Z.K. {Berta}, D.~{Charbonneau}, D.W. {Latham}, C.J.
  {Burke}, G.A. {Esquerdo}, M.E. {Everett}, M.J. {Holman}, P.~{Nutzman},
  P.~{Berlind}, M.L. {Calkins}, E.E. {Falco}, J.N. {Winn}, J.A. {Johnson}, J.Z.
  {Gazak}, ApJ \textbf{718}, 1353 (2010).
\newblock \doi{10.1088/0004-637X/718/2/1353}

\bibitem{deleuil08}
M.~{Deleuil}, H.J. {Deeg}, R.~{Alonso}, F.~{Bouchy}, D.~{Rouan}, M.~{Auvergne},
  A.~{Baglin}, S.~{Aigrain}, J.M. {Almenara}, M.~{Barbieri}, P.~{Barge},
  H.~{Bruntt}, P.~{Bord{\'e}}, A.~{Collier Cameron}, S.~{Csizmadia}, R.~{de La
  Reza}, R.~{Dvorak}, A.~{Erikson}, M.~{Fridlund}, D.~{Gandolfi}, M.~{Gillon},
  E.~{Guenther}, T.~{Guillot}, A.~{Hatzes}, G.~{H{\'e}brard}, L.~{Jorda},
  H.~{Lammer}, A.~{L{\'e}ger}, A.~{Llebaria}, B.~{Loeillet}, M.~{Mayor},
  T.~{Mazeh}, C.~{Moutou}, M.~{Ollivier}, M.~{P{\"a}tzold}, F.~{Pont},
  D.~{Queloz}, H.~{Rauer}, J.~{Schneider}, A.~{Shporer}, G.~{Wuchterl},
  S.~{Zucker}, A\&A \textbf{491}, 889 (2008).
\newblock \doi{10.1051/0004-6361:200810625}

\bibitem{gorlova03}
N.I. {Gorlova}, M.R. {Meyer}, G.H. {Rieke}, J.~{Liebert}, ApJ \textbf{593},
  1074 (2003)

\bibitem{gagne14a}
J.~{Gagn{\'e}}, D.~{Lafreni{\`e}re}, R.~{Doyon}, L.~{Malo}, {\'E}.~{Artigau},
  ApJ \textbf{783}, 121 (2014).
\newblock \doi{10.1088/0004-637X/783/2/121}

\bibitem{filippazzo15}
J.C. {Filippazzo}, E.L. {Rice}, J.~{Faherty}, K.L. {Cruz}, M.M. {Van Gordon},
  D.L. {Looper}, ApJ \textbf{810}, 158 (2015).
\newblock \doi{10.1088/0004-637X/810/2/158}

\bibitem{martin17a}
E.C. {Martin}, G.N. {Mace}, I.S. {McLean}, S.E. {Logsdon}, E.L. {Rice}, J.D.
  {Kirkpatrick}, A.J. {Burgasser}, M.R. {McGovern}, L.~{Prato}, ApJ  (2017)

\bibitem{lodieu18a}
N.~{Lodieu}, M.R. {Zapatero Osorio}, V.J.S. {B{\'e}jar}, K.~{Pe{\~n}a
  Ram{\'{\i}}rez}, Monthly Notices of the Royal Astronomical Society
  \textbf{473}, 2020 (2018).
\newblock \doi{10.1093/{Monthly Notices of the Royal Astronomical
  Society}/stx2279}

\bibitem{santos05a}
N.C. {Santos}, G.~{Israelian}, M.~{Mayor}, J.P. {Bento}, P.C. {Almeida}, S.G.
  {Sousa}, A.~{Ecuvillon}, A\&A \textbf{437}, 1127 (2005).
\newblock \doi{10.1051/0004-6361:20052895}

\bibitem{bond06}
J.C. {Bond}, C.G. {Tinney}, R.P. {Butler}, H.R.A. {Jones}, G.W. {Marcy}, A.J.
  {Penny}, B.D. {Carter}, Monthly Notices of the Royal Astronomical Society
  \textbf{370}, 163 (2006).
\newblock \doi{10.1111/j.1365-2966.2006.10459.x}

\bibitem{santos03a}
N.C. {Santos}, G.~{Israelian}, M.~{Mayor}, R.~{Rebolo}, S.~{Udry}, A\&A
  \textbf{398}, 363 (2003).
\newblock \doi{10.1051/0004-6361:20021637}

\bibitem{pinotti05}
R.~{Pinotti}, L.~{Arany-Prado}, W.~{Lyra}, G.F. {Porto de Mello}, Monthly
  Notices of the Royal Astronomical Society \textbf{364}, 29 (2005).
\newblock \doi{10.1111/j.1365-2966.2005.09491.x}

\bibitem{El_Badry19a}
K.~{El-Badry}, H.W. {Rix}, Monthly Notices of the Royal Astronomical Society
  \textbf{482}, L139 (2019).
\newblock \doi{10.1093/{Monthly Notices of the Royal Astronomical
  Society}l/sly206}

\bibitem{gizis99}
J.E. {Gizis}, I.N. {Reid}, AJ \textbf{117}, 508 (1999)

\bibitem{lepine07c}
S.~{L{\'e}pine}, R.M. {Rich}, M.M. {Shara}, ApJ \textbf{669}, 1235 (2007).
\newblock \doi{10.1086/521614}

\bibitem{burgasser07b}
A.J. {Burgasser}, K.L. {Cruz}, J.D. {Kirkpatrick}, ApJ \textbf{657}, 494
  (2007).
\newblock \doi{10.1086/510148}

\bibitem{jao08}
W.C. {Jao}, T.J. {Henry}, T.D. {Beaulieu}, J.P. {Subasavage}, AJ \textbf{136},
  840 (2008).
\newblock \doi{10.1088/0004-6256/136/2/840}

\bibitem{kirkpatrick16}
J.D. {Kirkpatrick}, K.~{Kellogg}, A.C. {Schneider}, S.~{Fajardo-Acosta}, M.C.
  {Cushing}, J.~{Greco}, G.N. {Mace}, C.R. {Gelino}, E.L. {Wright}, P.R.M.
  {Eisenhardt}, D.~{Stern}, J.K. {Faherty}, S.S. {Sheppard}, G.B. {Lansbury},
  S.E. {Logsdon}, E.C. {Martin}, I.S. {McLean}, S.D. {Schurr}, R.M. {Cutri},
  T.~{Conrow}, ApJS \textbf{224}, 36 (2016).
\newblock \doi{10.3847/0067-0049/224/2/36}

\bibitem{zhang17a}
Z.H. {Zhang}, D.J. {Pinfield}, M.C. {G{\'a}lvez-Ortiz}, B.~{Burningham},
  N.~{Lodieu}, F.~{Marocco}, A.J. {Burgasser}, A.C. {Day-Jones}, F.~{Allard},
  H.R.A. {Jones}, D.~{Homeier}, J.~{Gomes}, R.L. {Smart}, Monthly Notices of
  the Royal Astronomical Society \textbf{464}, 3040 (2017).
\newblock \doi{10.1093/{Monthly Notices of the Royal Astronomical
  Society}/stw2438}

\bibitem{moutou13}
C.~{Moutou}, A.S. {Bonomo}, G.~{Bruno}, G.~{Montagnier}, F.~{Bouchy}, J.M.
  {Almenara}, S.C.C. {Barros}, M.~{Deleuil}, R.F. {D{\'{\i}}az},
  G.~{H{\'e}brard}, A.~{Santerne}, A\&A \textbf{558}, L6 (2013).
\newblock \doi{10.1051/0004-6361/201322201}

\bibitem{Phillips1978}
M.J. {Phillips}, L.~{Hartmann}, ApJ \textbf{224}, 182 (1978).
\newblock \doi{10.1086/156363}

\bibitem{Wilson1978}
O.C. {Wilson}, ApJ \textbf{226}, 379 (1978).
\newblock \doi{10.1086/156618}

\bibitem{Baliunas1996}
S.L. {Baliunas}, E.~{Nesme-Ribes}, D.~{Sokoloff}, W.H. {Soon}, ApJ
  \textbf{460}, 848 (1996).
\newblock \doi{10.1086/177014}

\bibitem{Lockwood2007}
G.W. {Lockwood}, B.A. {Skiff}, G.W. {Henry}, S.~{Henry}, R.R. {Radick}, S.L.
  {Baliunas}, R.A. {Donahue}, W.~{Soon}, ApJS \textbf{171}, 260 (2007).
\newblock \doi{10.1086/516752}

\bibitem{Messina2002}
S.~{Messina}, E.F. {Guinan}, A\&A \textbf{393}, 225 (2002).
\newblock \doi{10.1051/0004-6361:20021000}

\bibitem{Olah2009}
K.~{Ol{\'a}h}, Z.~{Koll{\'a}th}, T.~{Granzer}, K.G. {Strassmeier}, A.F.
  {Lanza}, S.~{J{\"a}rvinen}, H.~{Korhonen}, S.L. {Baliunas}, W.~{Soon},
  S.~{Messina}, G.~{Cutispoto}, A\&A \textbf{501}, 703 (2009).
\newblock \doi{10.1051/0004-6361/200811304}

\bibitem{Jetsu1993}
L.~{Jetsu}, J.~{Pelt}, I.~{Tuominen}, A\&A \textbf{278}, 449 (1993)

\bibitem{Parihar2009}
P.~{Parihar}, S.~{Messina}, P.~{Bama}, B.J. {Medhi}, S.~{Muneer}, C.~{Velu},
  A.~{Ahmad}, Monthly Notices of the Royal Astronomical Society \textbf{395},
  593 (2009).
\newblock \doi{10.1111/j.1365-2966.2009.14422.x}

\bibitem{Eker2008}
Z.~{Eker}, N.F. {Ak}, S.~{Bilir}, D.~{Do{\v g}ru}, M.~{T{\"u}ys{\"u}z},
  E.~{Soydugan}, H.~{Bak{\i}{\c s}}, B.~{U{\v g}ra{\c s}}, F.~{Soydugan},
  A.~{Erdem}, O.~{Demircan}, Monthly Notices of the Royal Astronomical Society
  \textbf{389}, 1722 (2008).
\newblock \doi{10.1111/j.1365-2966.2008.13670.x}

\bibitem{Watson2004}
C.A. {Watson}, V.S. {Dhillon}, Monthly Notices of the Royal Astronomical
  Society \textbf{351}, 110 (2004).
\newblock \doi{10.1111/j.1365-2966.2004.07763.x}

\bibitem{Czesla2019}
S.~{Czesla}, S.~{Terzenbach}, R.~{Wichmann}, J.H.M.M. {Schmitt}, A\&A
  \textbf{623}, A107 (2019).
\newblock \doi{10.1051/0004-6361/201834516}

\bibitem{osten2016}
R.A. {Osten}, A.~{Kowalski}, S.A. {Drake}, H.~{Krimm}, K.~{Page}, K.~{Gazeas},
  J.~{Kennea}, S.~{Oates}, M.~{Page}, E.~{de Miguel}, R.~{Nov{\'a}k},
  T.~{Apeltauer}, N.~{Gehrels}, ApJ \textbf{832}, 174 (2016).
\newblock \doi{10.3847/0004-637X/832/2/174}

\bibitem{davenport2016}
J.R.A. {Davenport}, ApJ \textbf{829}(1), 23 (2016).
\newblock \doi{10.3847/0004-637X/829/1/23}

\bibitem{Ilin2019}
E.~{Ilin}, S.J. {Schmidt}, J.R.A. {Davenport}, K.G. {Strassmeier}, A\&A
  \textbf{622}, A133 (2019).
\newblock \doi{10.1051/0004-6361/201834400}

\bibitem{vandoor2017}
T.~{Van Doorsselaere}, H.~{Shariati}, J.~{Debosscher}, ApJS \textbf{232}(2), 26
  (2017).
\newblock \doi{10.3847/1538-4365/aa8f9a}

\bibitem{hawley2014}
S.L. {Hawley}, J.R.A. {Davenport}, A.F. {Kowalski}, J.P. {Wisniewski},
  L.~{Hebb}, R.~{Deitrick}, E.J. {Hilton}, ApJ \textbf{797}(2), 121 (2014).
\newblock \doi{10.1088/0004-637X/797/2/121}

\bibitem{gunther2019}
M.N. {G{\"u}nther}, Z.~{Zhan}, S.~{Seager}, P.B. {Rimmer}, S.~{Ranjan}, K.G.
  {Stassun}, R.J. {Oelkers}, T.~{Daylan}, E.~{Newton}, E.~{Gillen}, arXiv
  e-prints arXiv:1901.00443 (2019)

\bibitem{jackman2019}
J.A.G. {Jackman}, P.J. {Wheatley}, C.E. {Pugh}, D.Y. {Kolotkov}, A.M.
  {Broomhall}, G.M. {Kennedy}, S.J. {Murphy}, R.~{Raddi}, M.R. {Burleigh}, S.L.
  {Casewell}, MNRAS \textbf{482}(4), 5553 (2019).
\newblock \doi{10.1093/mnras/sty3036}

\bibitem{schmidt2019}
S.J. {Schmidt}, B.J. {Shappee}, J.L. {van Saders}, K.Z. {Stanek}, J.S. {Brown},
  C.S. {Kochanek}, S.~{Dong}, M.R. {Drout}, S.~{Frank}, T.W.S. {Holoien}, ApJ
  \textbf{876}(2), 115 (2019).
\newblock \doi{10.3847/1538-4357/ab148d}

\bibitem{chang2017}
H.Y. {Chang}, Y.H. {Song}, A.L. {Luo}, L.C. {Huang}, W.H. {Ip}, J.N. {Fu},
  Y.~{Zhang}, Y.H. {Hou}, Z.H. {Cao}, Y.F. {Wang}, ApJ \textbf{834}(1), 92
  (2017).
\newblock \doi{10.3847/1538-4357/834/1/92}

\bibitem{moral2009}
J.C. {Morales}, I.~{Ribas}, C.~{Jordi}, G.~{Torres}, J.~{Gallardo}, E.F.
  {Guinan}, D.~{Charbonneau}, M.~{Wolf}, D.W. {Latham},
  G.~{Anglada-Escud{\'e}}, ApJ \textbf{691}(2), 1400 (2009).
\newblock \doi{10.1088/0004-637X/691/2/1400}

\bibitem{torib2002}
G.~{Torres}, I.~{Ribas}, ApJ \textbf{567}(2), 1140 (2002).
\newblock \doi{10.1086/338587}

\bibitem{ribas2003}
I.~{Ribas}, A\&A \textbf{398}, 239 (2003).
\newblock \doi{10.1051/0004-6361:20021609}

\bibitem{lopez2005}
M.~{L{\'o}pez-Morales}, I.~{Ribas}, ApJ \textbf{631}(2), 1120 (2005).
\newblock \doi{10.1086/432680}

\bibitem{torres2013}
G.~{Torres}, Astronomische Nachrichten \textbf{334}(1-2), 4 (2013).
\newblock \doi{10.1002/asna.201211743}

\bibitem{feiden2014}
G.A. {Feiden}, B.~{Chaboyer}, A\&A \textbf{571}, A70 (2014).
\newblock \doi{10.1051/0004-6361/201424288}

\bibitem{ricker15}
G.R. {Ricker}, J.N. {Winn}, R.~{Vanderspek}, D.W. {Latham}, G.{\'A}. {Bakos},
  J.L. {Bean}, Z.K. {Berta-Thompson}, T.M. {Brown}, L.~{Buchhave}, N.R.
  {Butler}, R.P. {Butler}, W.J. {Chaplin}, D.~{Charbonneau},
  J.~{Christensen-Dalsgaard}, M.~{Clampin}, D.~{Deming}, J.~{Doty}, N.~{De
  Lee}, C.~{Dressing}, E.W. {Dunham}, M.~{Endl}, F.~{Fressin}, J.~{Ge},
  T.~{Henning}, M.J. {Holman}, A.W. {Howard}, S.~{Ida}, J.M. {Jenkins},
  G.~{Jernigan}, J.A. {Johnson}, L.~{Kaltenegger}, N.~{Kawai}, H.~{Kjeldsen},
  G.~{Laughlin}, A.M. {Levine}, D.~{Lin}, J.J. {Lissauer}, P.~{MacQueen},
  G.~{Marcy}, P.R. {McCullough}, T.D. {Morton}, N.~{Narita}, M.~{Paegert},
  E.~{Palle}, F.~{Pepe}, J.~{Pepper}, A.~{Quirrenbach}, S.A. {Rinehart},
  D.~{Sasselov}, B.~{Sato}, S.~{Seager}, A.~{Sozzetti}, K.G. {Stassun},
  P.~{Sullivan}, A.~{Szentgyorgyi}, G.~{Torres}, S.~{Udry}, J.~{Villasenor},
  Journal of Astronomical Telescopes, Instruments, and Systems \textbf{1}(1),
  014003 (2015).
\newblock \doi{10.1117/1.JATIS.1.1.014003}

\bibitem{roxburgh07}
I.~{Roxburgh}, C.~{Catala}, {PLATO Consortium}, Communications in
  Asteroseismology \textbf{150}, 357 (2007).
\newblock \doi{10.1553/cia150s357}

\bibitem{raetz16}
S.~{Raetz}, T.O.B. {Schmidt}, S.~{Czesla}, T.~{Klocov{\'a}}, L.~{Holmes},
  R.~{Errmann}, M.~{Kitze}, M.~{Fern{\'a}ndez}, A.~{Sota}, C.~{Brice{\~n}o},
  J.~{Hern{\'a}ndez}, J.J. {Downes}, D.P. {Dimitrov}, D.~{Kjurkchieva},
  V.~{Radeva}, Z.Y. {Wu}, X.~{Zhou}, H.~{Takahashi}, T.~{Henych},
  M.~{Seeliger}, M.~{Mugrauer}, C.~{Adam}, C.~{Marka}, J.G. {Schmidt}, M.M.
  {Hohle}, C.~{Ginski}, T.~{Pribulla}, L.~{Trepl}, M.~{Moualla}, N.~{Pawellek},
  J.~{Gelszinnis}, S.~{Buder}, S.~{Masda}, G.~{Maciejewski},
  R.~{Neuh{\"a}user}, Monthly Notices of the Royal Astronomical Society
  \textbf{460}, 2834 (2016).
\newblock \doi{10.1093/{Monthly Notices of the Royal Astronomical
  Society}/stw1159}

\bibitem{errmann13}
R.~{Errmann}, R.~{Neuh{\"a}user}, L.~{Marschall}, G.~{Torres}, M.~{Mugrauer},
  W.P. {Chen}, S.C.L. {Hu}, C.~{Briceno}, R.~{Chini}, {\L}.~{Bukowiecki}, D.P.
  {Dimitrov}, D.~{Kjurkchieva}, E.L.N. {Jensen}, D.H. {Cohen}, Z.Y. {Wu},
  T.~{Pribulla}, M.~{Va{\v n}ko}, V.~{Krushevska}, J.~{Budaj}, Y.~{Oasa}, A.K.
  {Pandey}, M.~{Fernandez}, A.~{Kellerer}, C.~{Marka}, Astronomische
  Nachrichten \textbf{334}, 673 (2013).
\newblock \doi{10.1002/asna.201311890}

\bibitem{garai16}
Z.~{Garai}, T.~{Pribulla}, {\v L}.~{Hamb{\'a}lek}, R.~{Errmann}, C.~{Adam},
  S.~{Buder}, T.~{Butterley}, V.S. {Dhillon}, B.~{Dincel}, H.~{Gilbert},
  C.~{Ginski}, L.K. {Hardy}, A.~{Kellerer}, M.~{Kitze}, E.~{Kundra}, S.P.
  {Littlefair}, M.~{Mugrauer}, J.~{Nedoro{\v s}{\v c}{\'{\i}}k},
  R.~{Neuh{\"a}user}, A.~{Pannicke}, S.~{Raetz}, J.G. {Schmidt}, T.O.B.
  {Schmidt}, M.~{Seeliger}, M.~{Va{\v n}ko}, R.W. {Wilson}, Astronomische
  Nachrichten \textbf{337}, 261 (2016).
\newblock \doi{10.1002/asna.201512310}

\bibitem{dobbie10}
P.D. {Dobbie}, N.~{Lodieu}, R.G. {Sharp}, Monthly Notices of the Royal
  Astronomical Society \textbf{409}, 1002 (2010).
\newblock \doi{10.1111/j.1365-2966.2010.17355.x}

\bibitem{manzi08}
S.~{Manzi}, S.~{Randich}, W.J. {de Wit}, F.~{Palla}, A\&A \textbf{479}, 141
  (2008).
\newblock \doi{10.1051/0004-6361:20078226}

\bibitem{barrado02a}
D.~{Barrado y Navascu{\'e}s}, J.~{Bouvier}, J.R. {Stauffer}, N.~{Lodieu}, M.J.
  {McCaughrean}, A\&A \textbf{395}, 813 (2002)

\bibitem{huensch04}
M.~{H{\"u}nsch}, S.~{Randich}, M.~{Hempel}, C.~{Weidner}, J.H.M.M. {Schmitt},
  A\&A \textbf{418}, 539 (2004).
\newblock \doi{10.1051/0004-6361:20040043}

\bibitem{emerson01}
J.P. {Emerson}, in \emph{The New Era of Wide Field Astronomy},
  \emph{Astronomical Society of the Pacific Conference Series}, vol. 232, ed.
  by {R.~Clowes, A.~Adamson, \& G.~Bromage} (2001), \emph{Astronomical Society
  of the Pacific Conference Series}, vol. 232, p. 339

\bibitem{dalton06}
G.B. {Dalton}, M.~{Caldwell}, A.K. {Ward}, M.S. {Whalley}, G.~{Woodhouse}, R.L.
  {Edeson}, P.~{Clark}, S.M. {Beard}, A.M. {Gallie}, S.P. {Todd}, J.M.D.
  {Strachan}, N.N. {Bezawada}, W.J. {Sutherland}, J.P. {Emerson}, in
  \emph{Society of Photo-Optical Instrumentation Engineers (SPIE) Conference
  Series}, \emph{Presented at the Society of Photo-Optical Instrumentation
  Engineers (SPIE) Conference}, vol. 6269 (2006), \emph{Presented at the
  Society of Photo-Optical Instrumentation Engineers (SPIE) Conference}, vol.
  6269.
\newblock \doi{10.1117/12.670018}

\bibitem{wheatley13}
P.J. {Wheatley}, D.L. {Pollacco}, D.~{Queloz}, H.~{Rauer}, C.A. {Watson}, R.G.
  {West}, B.~{Chazelas}, T.M. {Louden}, S.~{Walker}, N.~{Bannister},
  J.~{Bento}, M.~{Burleigh}, J.~{Cabrera}, P.~{Eigm{\"u}ller}, A.~{Erikson},
  L.~{Genolet}, M.~{Goad}, A.~{Grange}, A.~{Jord{\'a}n}, K.~{Lawrie},
  J.~{McCormac}, M.~{Neveu}, in \emph{European Physical Journal Web of
  Conferences}, \emph{European Physical Journal Web of Conferences}, vol.~47
  (2013), \emph{European Physical Journal Web of Conferences}, vol.~47, p.
  13002.
\newblock \doi{10.1051/epjconf/20134713002}

\bibitem{gillon11a}
M.~{Gillon}, E.~{Jehin}, P.~{Magain}, V.~{Chantry}, D.~{Hutsem{\'e}kers},
  J.~{Manfroid}, D.~{Queloz}, S.~{Udry}, in \emph{European Physical Journal Web
  of Conferences}, \emph{European Physical Journal Web of Conferences}, vol.~11
  (2011), \emph{European Physical Journal Web of Conferences}, vol.~11, p.
  06002.
\newblock \doi{10.1051/epjconf/20101106002}

\bibitem{gillon13a}
M.~{Gillon}, E.~{Jehin}, L.~{Delrez}, P.~{Magain}, C.~{Opitom}, S.~{Sohy}, in
  \emph{Protostars and Planets VI Posters} (2013)

\bibitem{alonso04a}
R.~{Alonso}, T.M. {Brown}, G.~{Torres}, D.W. {Latham}, A.~{Sozzetti},
  G.~{Mandushev}, J.A. {Belmonte}, D.~{Charbonneau}, H.J. {Deeg}, E.W.
  {Dunham}, F.T. {O'Donovan}, R.P. {Stefanik}, ApJL \textbf{613}, L153 (2004).
\newblock \doi{10.1086/425256}

\bibitem{bakos02}
G.{\'A}. {Bakos}, J.~{L{\'a}z{\'a}r}, I.~{Papp}, P.~{S{\'a}ri}, E.M. {Green},
  Publications of the Astronomical Society of the Pacific \textbf{114}, 974
  (2002).
\newblock \doi{10.1086/342382}

\bibitem{pollacco06}
D.L. {Pollacco}, I.~{Skillen}, A.~{Collier Cameron}, D.J. {Christian},
  C.~{Hellier}, J.~{Irwin}, T.A. {Lister}, R.A. {Street}, R.G. {West}, D.R.
  {Anderson}, W.I. {Clarkson}, H.~{Deeg}, B.~{Enoch}, A.~{Evans},
  A.~{Fitzsimmons}, C.A. {Haswell}, S.~{Hodgkin}, K.~{Horne}, S.R. {Kane}, F.P.
  {Keenan}, P.F.L. {Maxted}, A.J. {Norton}, J.~{Osborne}, N.R. {Parley}, R.S.I.
  {Ryans}, B.~{Smalley}, P.J. {Wheatley}, D.M. {Wilson}, Publications of the
  Astronomical Society of the Pacific \textbf{118}, 1407 (2006).
\newblock \doi{10.1086/508556}

\bibitem{mccullough05a}
P.R. {McCullough}, J.E. {Stys}, J.A. {Valenti}, S.W. {Fleming}, K.A. {Janes},
  J.N. {Heasley}, Publications of the Astronomical Society of the Pacific
  \textbf{117}, 783 (2005).
\newblock \doi{10.1086/432024}

\bibitem{pepper07}
J.~{Pepper}, R.W. {Pogge}, D.L. {DePoy}, J.L. {Marshall}, K.Z. {Stanek}, A.M.
  {Stutz}, S.~{Poindexter}, R.~{Siverd}, T.P. {O'Brien}, M.~{Trueblood},
  P.~{Trueblood}, Publications of the Astronomical Society of the Pacific
  \textbf{119}, 923 (2007).
\newblock \doi{10.1086/521836}

\bibitem{montes01b}
D.~{Montes}, J.~{L{\'o}pez-Santiago}, M.J. {Fern{\'a}ndez-Figueroa}, M.C.
  {G{\'a}lvez}, A\&A \textbf{379}, 976 (2001).
\newblock \doi{10.1051/0004-6361:20011385}

\bibitem{zuckerman04}
B.~{Zuckerman}, I.~{Song}, ARA\&A \textbf{42}, 685 (2004)

\bibitem{antoja08}
T.~{Antoja}, F.~{Figueras}, D.~{Fern{\'a}ndez}, J.~{Torra}, A\&A \textbf{490},
  135 (2008).
\newblock \doi{10.1051/0004-6361:200809519}

\bibitem{Hubrig2007}
S.~{Hubrig}, O.~{Marco}, B.~{Stelzer}, M.~{Sch{\"o}ller}, N.~{Hu{\'e}lamo},
  Monthly Notices of the Royal Astronomical Society \textbf{381}, 1569 (2007).
\newblock \doi{10.1111/j.1365-2966.2007.12325.x}

\bibitem{Evans2018}
I.N. {Evans}, F.~{Civano}, Astronomy and Geophysics \textbf{59}(2), 2.17
  (2018).
\newblock \doi{10.1093/astrogeo/aty079}

\bibitem{Boller2016}
T.~{Boller}, M.J. {Freyberg}, J.~{Tr{\"u}mper}, F.~{Haberl}, W.~{Voges},
  K.~{Nandra}, A\&A \textbf{588}, A103 (2016).
\newblock \doi{10.1051/0004-6361/201525648}

\bibitem{Traulsen2019}
I.~{Traulsen}, A.D. {Schwope}, G.~{Lamer}, J.~{Ballet}, F.~{Carrera},
  M.~{Coriat}, M.J. {Freyberg}, L.~{Michel}, C.~{Motch}, S.R. {Rosen},
  N.~{Webb}, M.T. {Ceballos}, F.~{Koliopanos}, J.~{Kurpas}, M.J. {Page}, M.G.
  {Watson}, A\&A \textbf{624}, A77 (2019).
\newblock \doi{10.1051/0004-6361/201833938}

\bibitem{Schroeder2007}
C.~{Schr{\"o}der}, J.H.M.M. {Schmitt}, A\&A \textbf{475}, 677 (2007).
\newblock \doi{10.1051/0004-6361:20077429}

\bibitem{Stelzer2013}
B.~{Stelzer}, A.~{Marino}, G.~{Micela}, J.~{L{\'o}pez-Santiago}, C.~{Liefke},
  Monthly Notices of the Royal Astronomical Society \textbf{431}, 2063 (2013).
\newblock \doi{10.1093/mnras/stt225}

\end{thebibliography}
%

%\begin{thebibliography}{99.}%
%% and use \bibitem to create references.
%%
%% Use the following syntax and markup for your references if 
%% the subject of your book is from the field 
%% "Mathematics, Physics, Statistics, Computer Science"
%%
%% Contribution 
%\bibitem{science-contrib} Broy, M.: Software engineering --- from auxiliary to key technologies. In: Broy, M., Dener, E. (eds.) Software Pioneers, pp. 10-13. Springer, Heidelberg (2002)
%%
%% Online Document
%\bibitem{science-online} Dod, J.: Effective substances. In: The Dictionary of Substances and Their Effects. Royal Society of Chemistry (1999) Available via DIALOG. \\
%\url{http://www.rsc.org/dose/title of subordinate document. Cited 15 Jan 1999}
%%
%% Monograph
%\bibitem{science-mono} Geddes, K.O., Czapor, S.R., Labahn, G.: Algorithms for Computer Algebra. Kluwer, Boston (1992) 
%%
%% Journal article
%\bibitem{science-journal} Hamburger, C.: Quasimonotonicity, regularity and duality for nonlinear systems of partial differential equations. Ann. Mat. Pura. Appl. \textbf{169}, 321--354 (1995)
%%
%% Journal article by DOI
%\bibitem{science-DOI} Slifka, M.K., Whitton, J.L.: Clinical implications of dysregulated cytokine production. J. Mol. Med. (2000) doi: 10.1007/s001090000086
%%
%\end{thebibliography}
\end{document}